\documentclass[sigconf]{acmart}

\AtBeginDocument{%
  }

\setcopyright{acmlicensed}
\copyrightyear{2025}
\acmYear{2025}
\acmDOI{XXXXXXX.XXXXXXX}

\acmConference[SIGMOD'25]{the 2025 ACM SIGMOD International Conference on Management of Data}{June 22--27, 2025}{Berlin, Germany}
\acmISBN{978-1-4503-XXXX-X/18/06}

\usepackage{algorithm}
\usepackage{algpseudocode}
\usepackage{amsmath}
\usepackage{subfig}
\usepackage{float}
\usepackage{url}
\usepackage{tabularx}
\usepackage{makecell}
\usepackage{tikz}
\newcommand*{\circled}[1]{\lower.6ex\hbox{\tikz\draw (0pt, 0pt) circle (.4em) node {\makebox[1em][c]{\small #1}};}}

\usepackage{listings}
\usepackage{xspace}
\usepackage{hyperref}
\usepackage[noabbrev, capitalize, nameinlink]{cleveref}

\newcommand{\sysname}{Accordion\xspace}
\newcommand{\presto}{Presto\xspace}
\newcommand{\prestissimo}{Prestissimo\xspace}
\newcommand{\arrow}{Apache Arrow\xspace}

\begin{document}

\title{Intra-Query Runtime Elasticity for Cloud-Native Data Analysis}


%%\author{Ben Trovato}
%%\authornote{Both authors contributed equally to this research.}
%%\email{trovato@corporation.com}
%%\orcid{1234-5678-9012}
%%\author{G.K.M. Tobin}
%%\authornotemark[1]
%%\email{webmaster@marysville-ohio.com}
%%\affiliation{%
%%  \institution{Institute for Clarity in Documentation}
%% \city{Dublin}
%%  \state{Ohio}
%%  \country{USA}
%%}

%\author{Xukang Zhang}
%\affiliation{%
%  \institution{Renmin University of China}
%\email{zhangxk@ruc.edu.cn}
%}

%\author{Huanchen Zhang}
%\affiliation{%
%  \institution{Tsinghua University}
%\email{huanchen@tsinghua.edu.cn}
%}

%\author{Xiaofeng Meng}
%\affiliation{%
% \institution{Renmin University of China}
%\email{xfmeng@ruc.edu.cn}
%}

\author{Xukang Zhang}
%\email{webmaster@marysville-ohio.com}
\affiliation{%
  \institution{Renmin University of China}
  \country{Beijing, China}
}
\email{zhangxk@ruc.edu.cn}

\author{Huanchen Zhang}
\authornote{Huanchen Zhang is also affiliated with the Shanghai Qi Zhi Institute.}
\affiliation{%
  \institution{Tsinghua University}
  %\city{Hekla}
  \country{Beijing, China}
}
\email{huanchen@tsinghua.edu.cn}

\author{Xiaofeng Meng}
\authornote{Xiaofeng Meng is the corresponding author.}
\affiliation{%
  \institution{Renmin University of China}
  %\city{Rocquencourt}
  \country{Beijing, China}
}
\email{xfmeng@ruc.edu.cn}

%%\author{Huifen Chan}
%%\affiliation{%
%%  \institution{Tsinghua University}
%%  \city{Haidian Qu}
%%  \state{Beijing Shi}
%%  \country{China}}

%%\author{Charles Palmer}
%%\affiliation{%
%%  \institution{Palmer Research Laboratories}
%%  \city{San Antonio}
%%  \state{Texas}
%%  \country{USA}}
%%\email{cpalmer@prl.com}

%%\author{John Smith}
%%\affiliation{%
%%  \institution{The Th{\o}rv{\"a}ld Group}
%%  \city{Hekla}
%%  \country{Iceland}}
%%\email{jsmith@affiliation.org}

%%\author{Julius P. Kumquat}
%%\affiliation{%
%%  \institution{The Kumquat Consortium}
%%  \city{New York}
%%  \country{USA}}
%%\email{jpkumquat@consortium.net}

%%\renewcommand{\shortauthors}{Trovato et al.}

\begin{abstract}

We propose the concept of \emph{Intra-Query Runtime Elasticity} (IQRE) for cloud-native data analysis. IQRE enables a cloud-native OLAP engine to dynamically adjust a query's \emph{Degree of Parallelism} (DOP) during execution. This capability allows users to utilize cloud computing resources more cost-effectively. We present \sysname, the first IQRE query engine. \sysname can adjust the parallelism of a query at any point during query execution without pausing data processing. It features a user-friendly interface and an auto-tuner backed by a ``what-if'' service to allow users to adjust the DOP according to their query latency constraints. The design of \sysname follows the execution model in \presto, an open-source distributed SQL query engine developed at Meta. We present the implementation of \sysname and demonstrate its ease of use, showcasing how it enables users to minimize compute resource consumption while meeting their query time constraints.

\end{abstract}

%%  \includegraphics[width=\textwidth]{sampleteaser}
%%  \caption{Seattle Mariners at Spring Training, 2010.}
%%  \Description{Enjoying the baseball game from the third-base
%%  seats. Ichiro Suzuki preparing to bat.}


\maketitle

\section{Introduction}

The emergence of cloud-native databases\cite{snowflakeWeb, redshiftWeb, bigQueryWeb, microsoftWeb} allows efficient data analysis in the cloud environment. Leveraging massively parallel processing engines \cite{prestoWeb, impalaWeb, veloxWeb}, these systems provide users with a robust parallel data processing experience, harnessing the extensive computational resources available in the cloud. Nonetheless, the challenge of economically using cloud databases remains inadequately addressed. Users often struggle to determine the optimal allocation of computing resources within their temporal and financial constraints, primarily due to the difficulty in predicting the relationship between resource utilization and query execution time before query execution. Existing methodologies \cite{userworkload1,userworkload2,userworkload3,userworkload4,userworkload5} typically involve constructing performance-cost models that necessitate the execution of specific user-provided workloads. These methods are time-consuming, less accessible for non-specialized users, and often lack generalizability \cite{costIntelligent}. However, the time-resource relationship is not available only after the query is executed. During query execution, by collecting runtime information (table scanning rate, throughput rate) of the query, it is possible to predict the relationship between the remaining time of query execution and resource usage (degree of parallelism). If parallelism can be dynamically adjusted during query execution, users could more effectively align execution time and resource expenditure with their budgetary requirements by the predicted relationship.

In this paper, we introduce the concept of \emph{Intra-Query Runtime Elasticity} (IQRE) and present the first IQRE query engine, named \textbf{\sysname}. IQRE refers to the capability of dynamically adjusting the parallelism of a query during execution without pausing data processing. This approach allows users to initiate a query with a minimal allocation of computational resources and subsequently modify the execution speed or resource consumption according to their requirements.

\begin{figure}
  \includegraphics[width=1\linewidth]{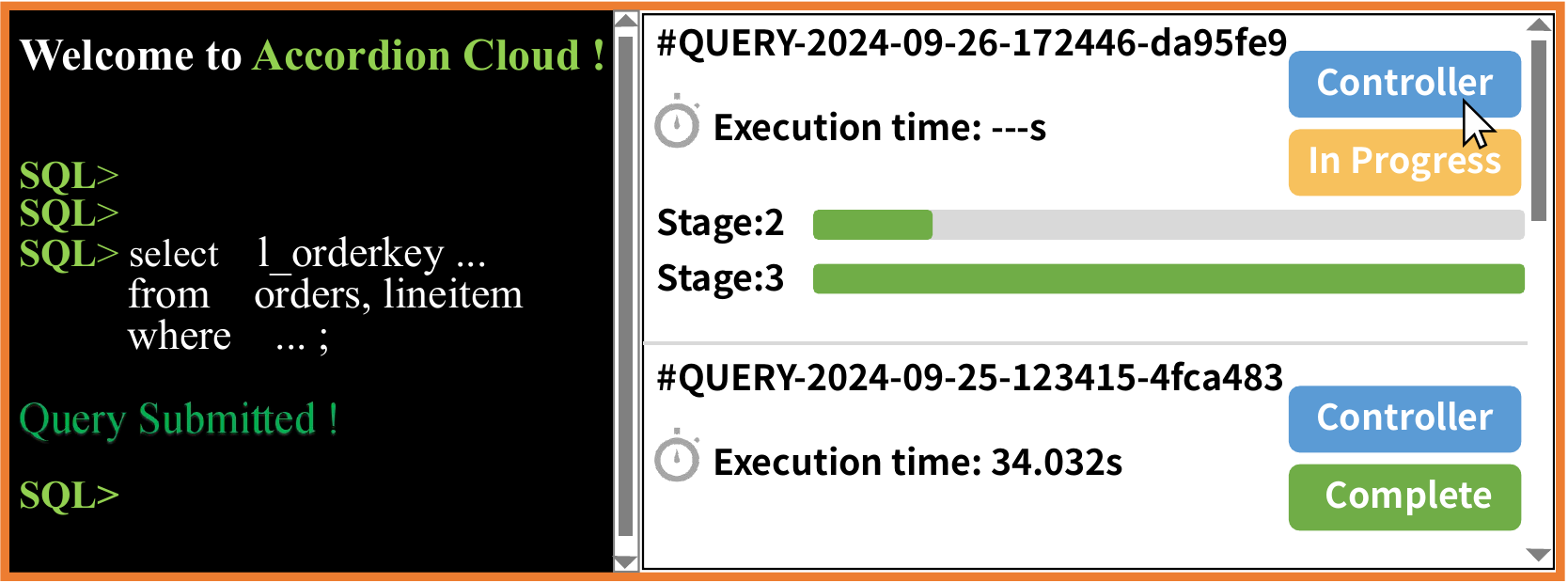}
  % \caption{The main interface is divided into two sections—the SQL input box on the left and the query execution progress tracking box on the right.}
  \caption{\sysname's Main Interface \textnormal{-- it includes a SQL input box on the left and the query execution progress tracking box on the right.}}
  \label{mainui}
\end{figure}

\sysname\footnote{\url{https://github.com/Blueratzxk/Accordion_engine}} was implemented in C++ from scratch, following the execution model in \presto\cite{presto-rbo}, an open-source distributed SQL engine developed by Meta.
%\revision{M3: R2.C1}{(the approach proposed in this paper mainly targets the vectorized and push-based query engines)}.
\sysname's execution engine adopts the vectorized push-based model
and uses \arrow\cite{arrowWeb} as the data exchange format between compute nodes.
\sysname features a user-friendly interface to facilitate adjusting the \emph{Degree of Parallelism} (DOP) at query execution time.
As shown in \cref{mainui}, users enter SQL statements in the query input box, which will be submitted to the \sysname cluster for execution.
Running queries are displayed in the query progress tracking box. Each query contains multiple progress bars (corresponding to different stages).
The query execution finishes when all the progress bars are filled.

\begin{figure}
  \includegraphics[width=1\linewidth]{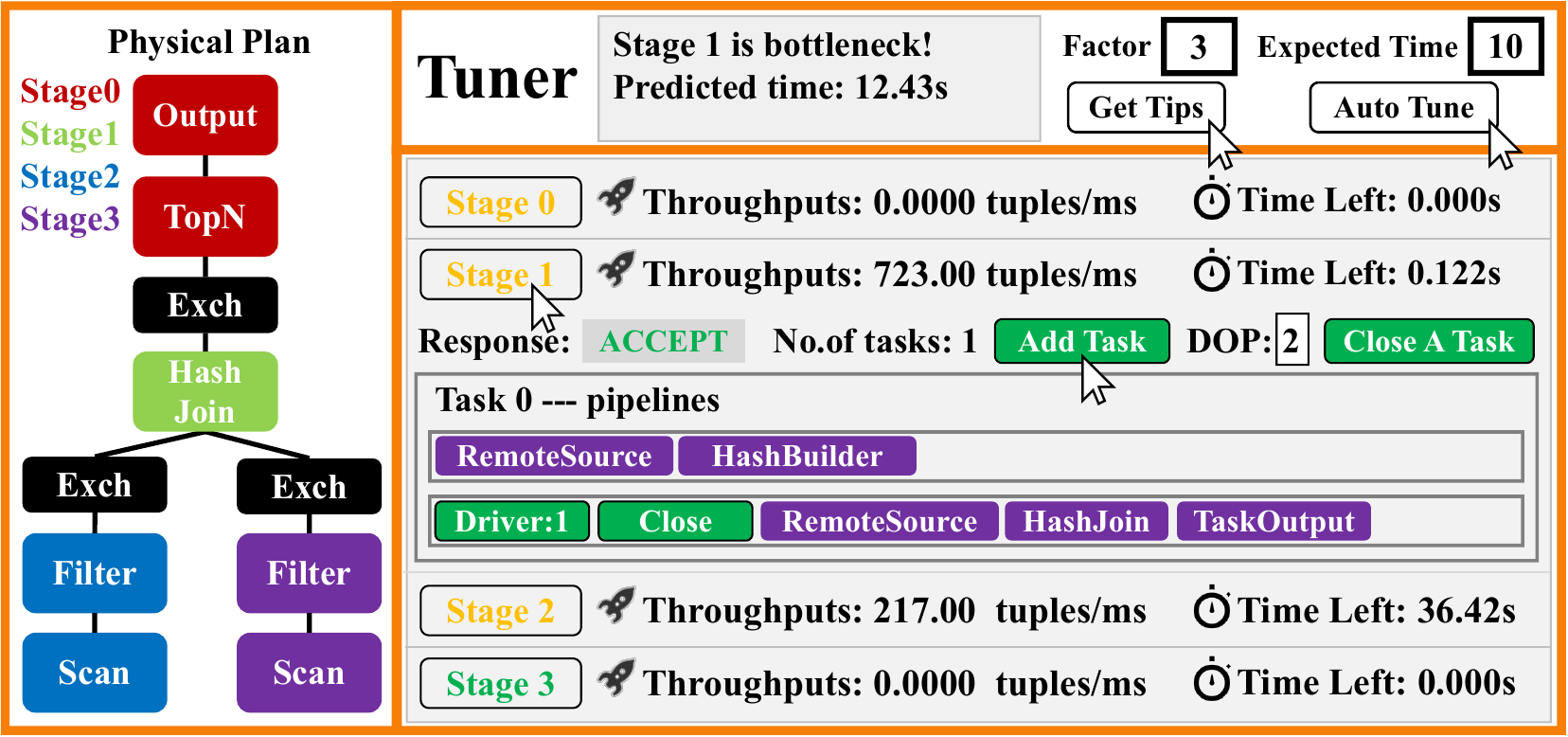}
  \caption{\sysname's Controller Interface \textnormal{-- it is composed of three sections: the query plan display box, the auto-tuner box, and the stage information box.}}
  \label{controller}
\end{figure}

Users can adjust the parallelism for each stage at execution time by tuning the DOP knobs in the controller interface (\cref{controller}). The controller interface provides detailed runtime information, including the query plan, real-time throughput for each stage, and the estimated remaining execution time. We also provide an auto-tuner backed by a ``what-if'' service that can help users automatically tune the query's DOP to meet their latency constraints.

This paper makes three primary contributions. First, we propose intra-query runtime elasticity (IQRE) for cloud-native databases as an important step toward cost-intelligent query processing. Second, we introduce \sysname, the first query engine that implements IQRE efficiently. Finally, we demonstrate that \sysname is easy to use and can use as few compute resources as possible to satisfy the query's latency constraint.

\section{Background}
\label{background}

\presto\cite{presto-rbo} has been widely used by enterprises and cloud database vendors for large-scale data analysis due to its high flexibility and elasticity. It is a query engine without storage components.
In this section, we provide an overview of \presto's architecture and discuss the challenges of implementing IQRE directly in Presto.

\begin{figure}[!t]
  \includegraphics[width=0.9\linewidth]{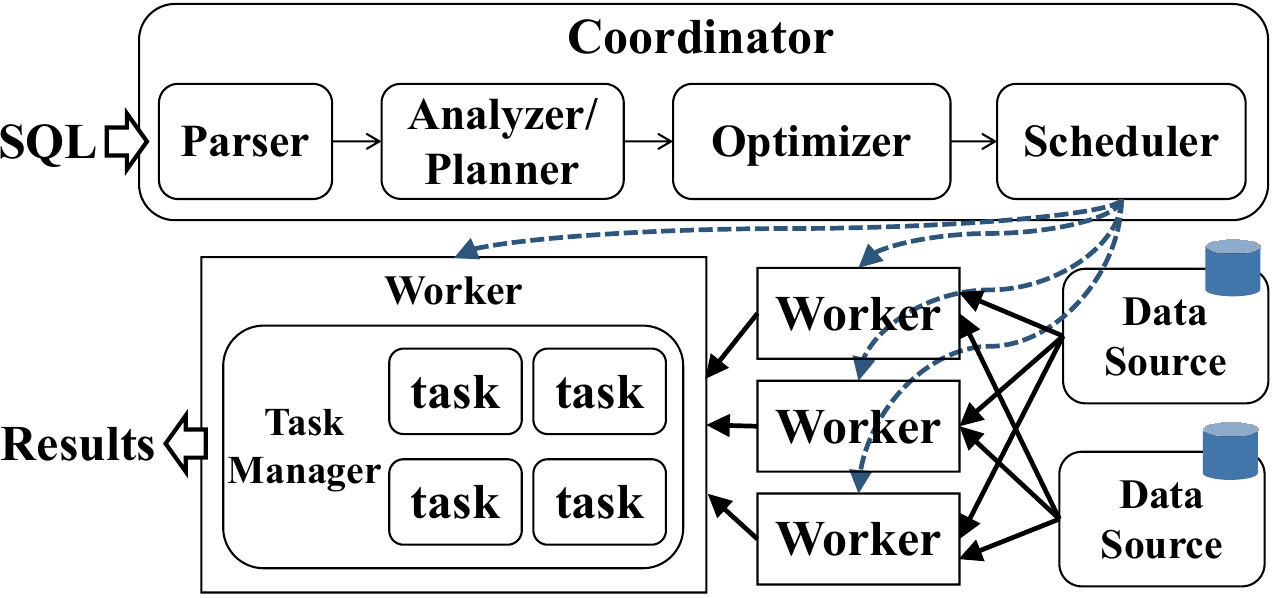}
  \caption{Architecture of \presto.}
  \label{PrestoArch}
\end{figure}

As illustrated in \cref{PrestoArch}, a \presto cluster consists of a coordinator node and multiple worker nodes. The coordinator is responsible for query parsing, analyzing, planning, optimizing, and task scheduling. Worker nodes are responsible for query processing and result return. Upon receiving a query, the coordinator analyzes the SQL statement, generates a distributed physical plan through optimization, and then schedules tasks — the smallest unit for distributed execution — on the worker nodes. Each worker node contains a task manager for creating and terminating tasks. Worker nodes execute these tasks to process data from base tables or to handle intermediate data generated by other workers. \presto uses RPC to exchange data between tasks.

% Data exchange between tasks occurs via RPC during processing.

%\begin{figure}
%  \includegraphics[width=0.7\linewidth]{pics/presto/logicalplan.pdf}
%  \caption{Logical plan of example query.}
%  \label{examquery}
%\end{figure}

\begin{figure}
  \includegraphics[width=0.8\linewidth]{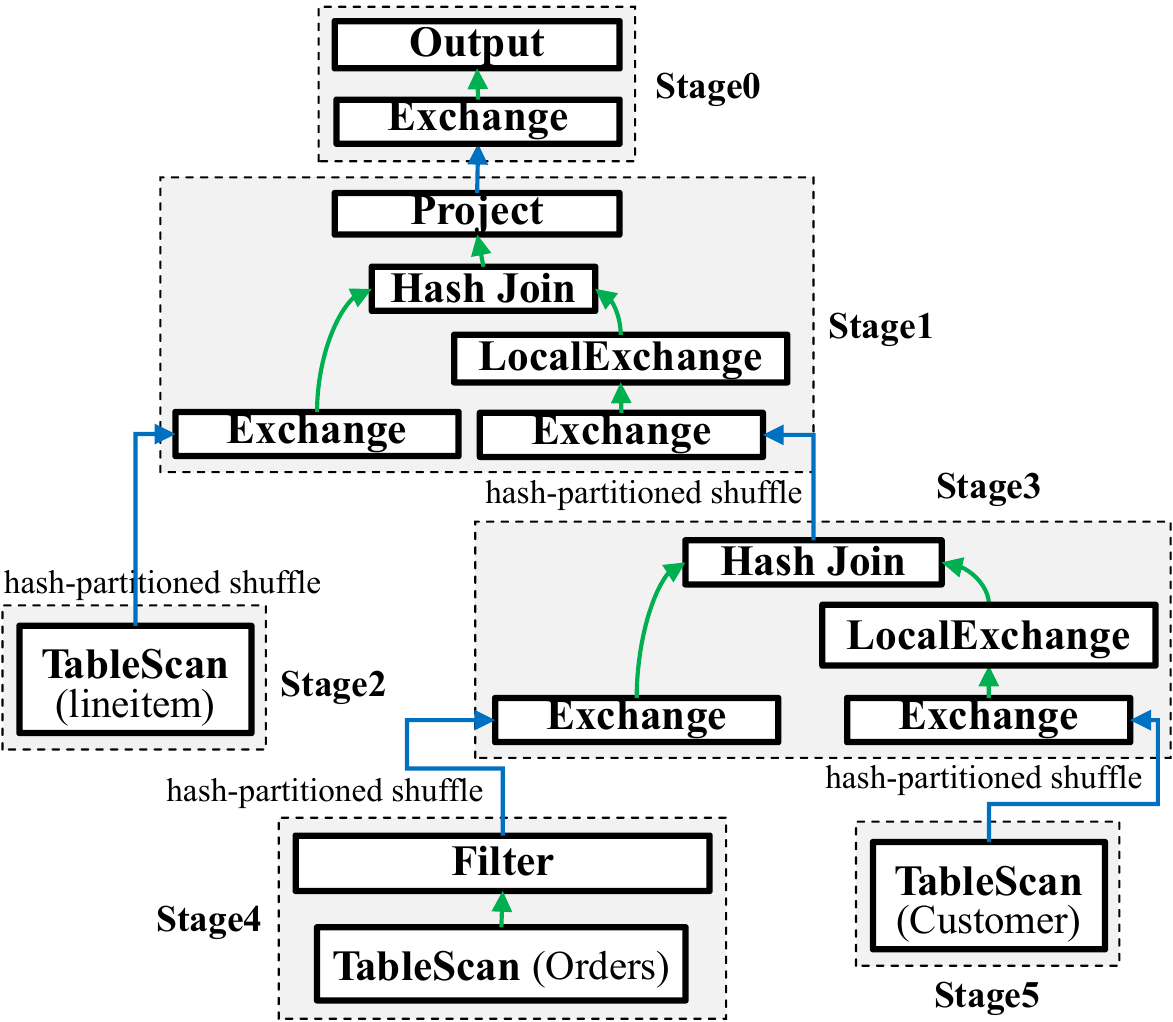}
  \caption{Distributed physical plan of example query. }
  \label{dexamquery}
\end{figure}

\begin{figure}
  \includegraphics[width=0.95\linewidth]{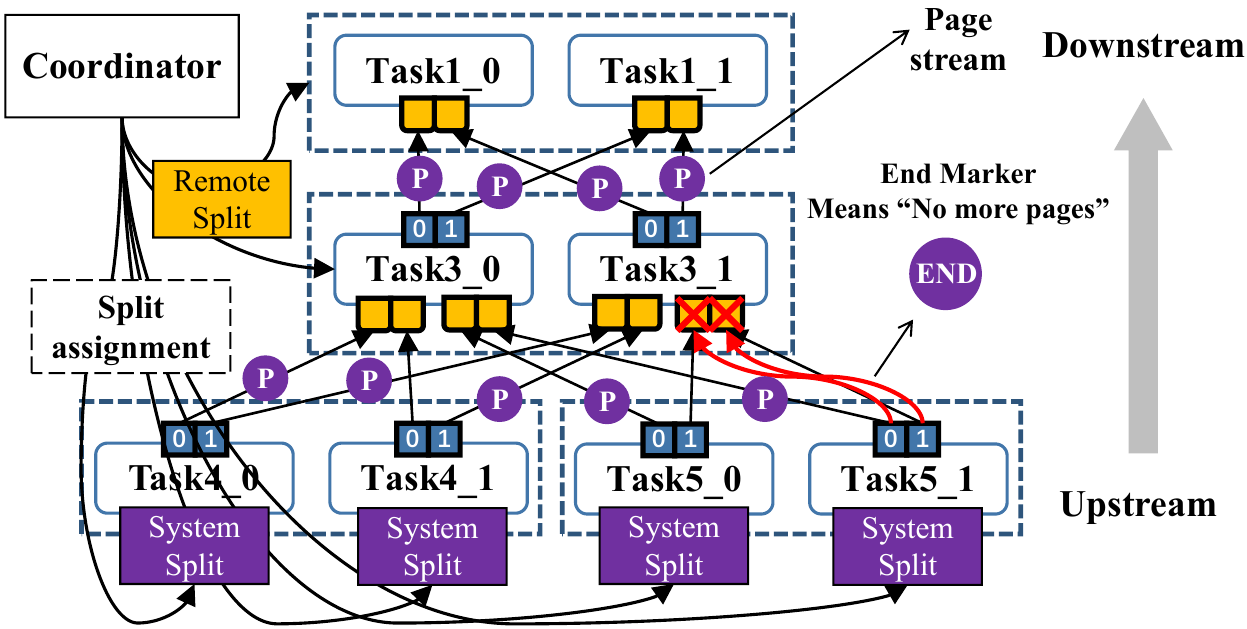}
  \caption{Partial distributed execution plan of the distributed physical plan. }
  \label{exexecutionplan}
\end{figure}

\vspace{0.5em}
\noindent\textbf{Physical Plan to Fragments.} Consider a simple query:

\begin{lstlisting}[language=SQL]
SELECT l_orderkey 
FROM lineitem 
    INNER JOIN orders  ON l_orderkey=o_orderkey
    INNER JOIN customer ON c_custkey=o_custkey
WHERE o_orderdate < 1994-03-05
\end{lstlisting}

\presto obtains a physical plan after parsing, analyzing, and optimizing the query. There are two special types of nodes in the plan: the exchange node and the local exchange node.
These nodes are introduced during query optimization to partition the plan into sub-plans.
The query optimizer divides the physical plan into multiple fragments based on the locations of the exchange nodes, resulting in a fragment (stage) tree as illustrated in \cref{dexamquery}. The scheduler allocates tasks across the cluster based on this stage tree to create a distributed execution plan. An execution stage includes multiple tasks. Each task is mapped to a compute node. \cref{exexecutionplan} presents a partially distributed execution plan for the stage tree, displaying only stages 1, 3, 4, and 5. Each stage is assigned two tasks, with each task identified by a unique task ID that consists of the stage number and the task sequence number.

\begin{figure}
  \includegraphics[width=1\linewidth]{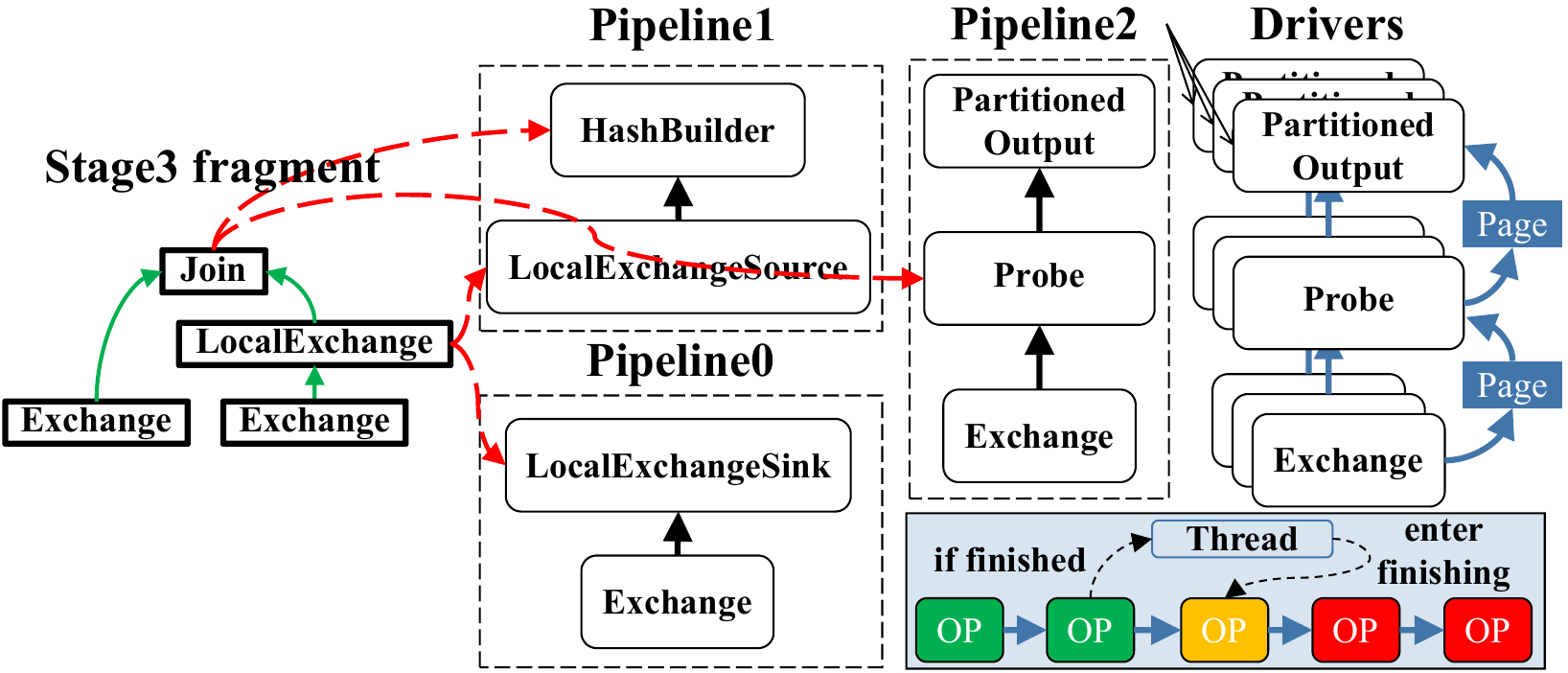}
  \caption{Fragment to pipelines \textnormal{-- fragment is divided into pipelines using pipeline breakers.}}
  \label{ftops}
\end{figure}

\vspace{0.5em}
\noindent\textbf{Fragment to Pipelines.} A fragment cannot be executed directly within a task; it must first be rewritten and then subdivided into a collection of pipelines. The division is performed by pipeline breakers, including the local exchange node and the hash join node in this plan. \cref{ftops} illustrates the process of converting a fragment into pipelines within the task of stage 3. Initially, the fragment is rewritten to introduce an output node. Subsequently, each local exchange node is divided into a sink node and a source node, while each join node is split into a probe node and a build node. This process results in a collection of plan node sequences, each of which will be transformed into a pipeline. A pipeline is defined as a sequence of operator factories, each capable of producing multiple physical operators. Consequently, each pipeline can generate physical operator sequences (driver), which represent the smallest unit of scheduling and execution in a task (the relationship between pipeline and driver is similar to the relationship between class and object in object-oriented programming).

\vspace{0.5em}
\noindent\textbf{Driver Execution.} Each driver can be executed by threads (the task manager keeps a thread pool and will spawn multiple drivers for each task; the drivers are scheduled by the task manager using a multi-level queue). Drivers involved in the table scan stage and those containing exchange operators require RPC addresses for execution. As depicted in \cref{exexecutionplan}, \presto utilizes the ``split'' object to set and update these addresses for drivers. There are primarily two types of splits in \presto: remote splits and system splits. A remote split, which includes a node's URL and a task ID, is used to establish data exchange connections between intermediate-stage (non-table scan stage) tasks and upstream stages' tasks. A system split is used to tell the table scan stage tasks where to get data chunks from external data sources for processing.

In the table scan stage, a data chunk is divided into smaller pages (sub-chunks), which are distributed among tasks for parallel processing. Pages also passed between physical operators. As illustrated in \cref{ftops}, each physical operator, driven by a thread, sequentially performs page input, processing, and output.

Each physical operator can exist in one of three states: finished, unfinished, or finishing. When a driver needs to be closed, the thread transitions each operator to the finished state in succession. Once all operators have reached the finished state, the driver is destroyed.

\begin{figure}
  \includegraphics[width=0.8\linewidth]{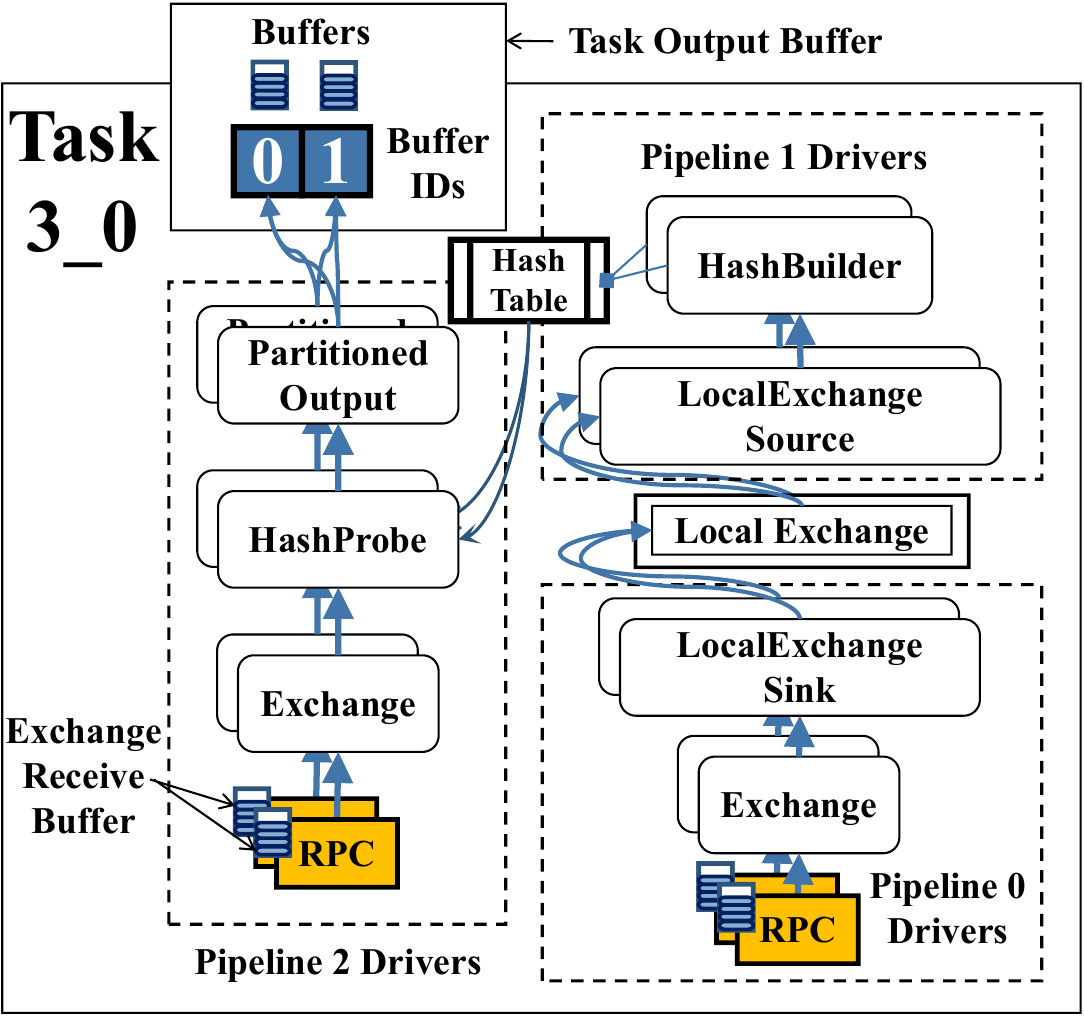}
  \caption{Internal details of stage 3's task.}
  \label{prestotask}
\end{figure}

\vspace{0.5em}
\noindent\textbf{Task Execution.} To illustrate the execution process of a task, we will use the task from stage 3 as an example. \cref{prestotask} details the execution of this task. Each pipeline generates two drivers. Pipeline 0 and Pipeline 2 request pages from the upstream tasks via exchange operators. Each exchange operator contains a receive buffer, which temporarily stores the data retrieved from upstream tasks. The driver of pipeline 0 passes the page to the local exchange structure (generated from the local exchange node) for hash partitioning. Pipeline 1 gets pages from the local exchange structure to build the hash table. Pipeline 2 receives the data and performs the probe operation. The probe result is hash-partitioned by the task output operator (containing a hash function) and then stored in the task output buffer. This buffer contains a vector of buffer IDs, each corresponding to a downstream task whose task sequence number matches the buffer ID. These downstream tasks then access pages using their task sequence numbers.

As shown in \cref{exexecutionplan}, if a task has no more pages to process, it will send ``end pages'' to notify the downstream tasks. With the help of the end page, the query's tasks can be automatically closed in a bottom-up fashion.

\vspace{0.5em}
\noindent\textbf{Challenges.} Implementing IQRE in \presto requires adjusting the number of tasks within a stage (stage DOP) or the number of drivers within a task (task DOP) during query execution. This is difficult because of the following challenges. First, \presto establishes the stage and task DOPs before query execution and does not permit modifications during query processing. It requires a dynamic scheduler capable of spawning or terminating tasks and drivers at runtime to break such early bindings. Second, the data exchange topology between tasks and drivers is fixed at query planning time in \presto. Modifying this topology requires extensive changes to various components, including the output buffers, drivers, task output operators, hash functions, etc. Third, \presto adopts a fixed capacity (configurable, default 32 MB) for the task output buffers. When the buffers are too large, tasks from the downstream stage might starve, waiting for data to process. This makes DOP tuning at this stage ineffective. On the other hand, if the buffers are too small, the network overhead becomes significant.

We will address these challenges in \cref{IQRESection}. First, we present the new architecture of \sysname in the next section.

\section{System Overview}
\label{overview}

\begin{figure}
  \includegraphics[scale=0.3]{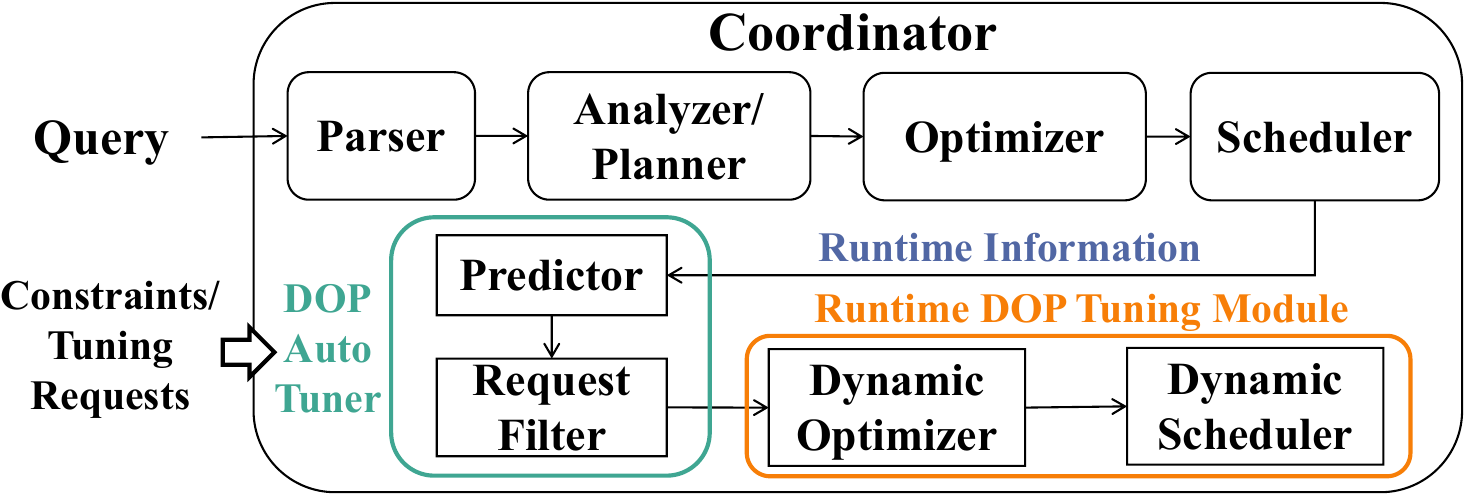}
  \caption{Architecture of \sysname.}
  \label{arch}
\end{figure}

\begin{figure}
  \includegraphics[scale=0.30]{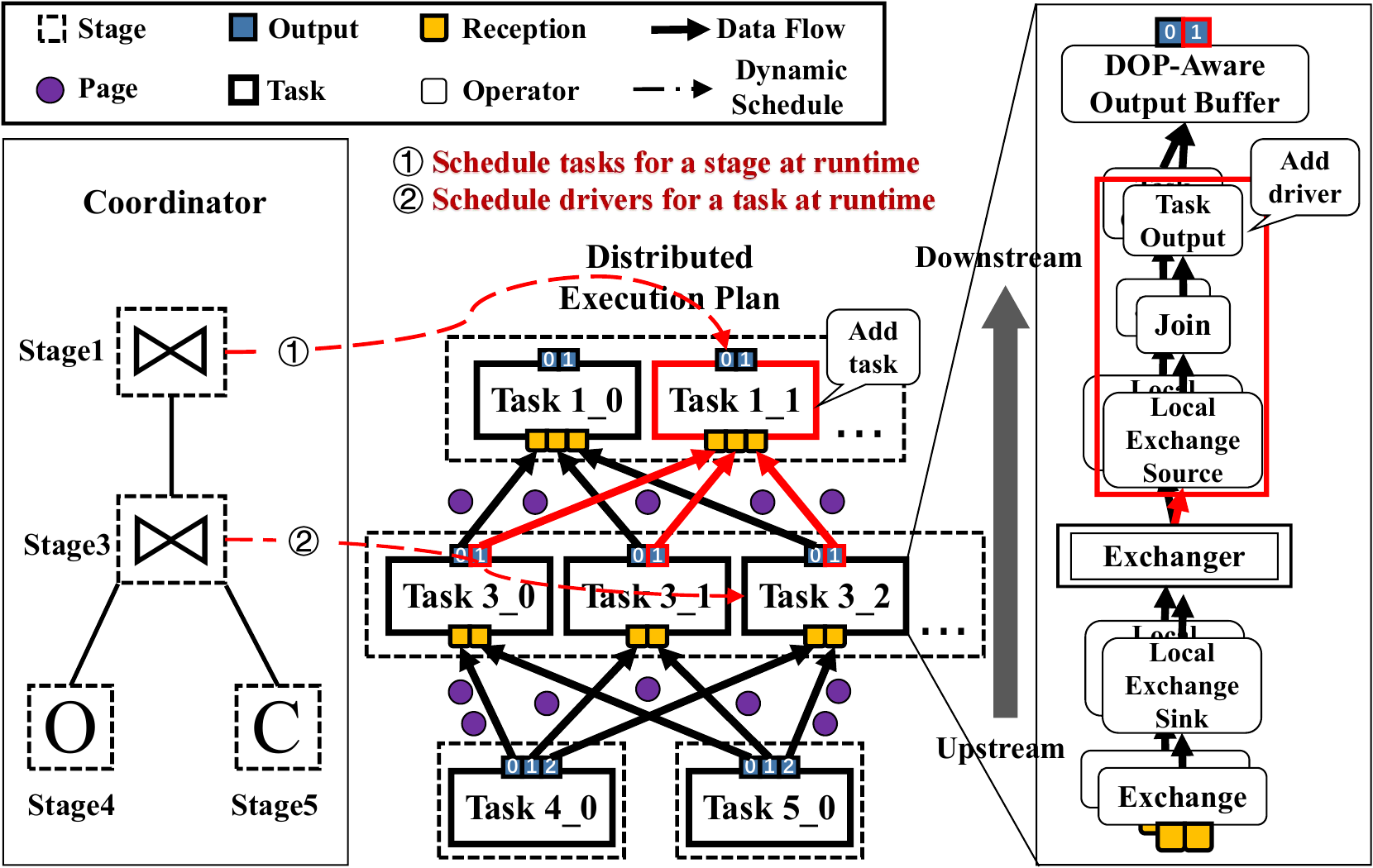}
  \caption{DOP tuning types in \sysname\textnormal{-- intra-task DOP runtime tuning (\textcircled{2}) and intra-stage DOP runtime tuning (\textcircled{1}). }}
  \label{chouxiang}
\end{figure}

Accordion is also a vectorized and push-based query engine like \presto. As shown in \cref{arch}, \sysname introduces a DOP auto-tuner and a runtime DOP tuning module on top of the existing design. The auto-tuner contains a predictor and tuning request filter. The Predictor (what-if service, as detailed in \cref{strategies}) handles prediction tasks. It obtains query runtime information from the scheduler to estimate the remaining execution time and the anticipated time after parallelism adjustments. These results are returned to users or used for DOP auto-tuning. The request filter is used to filter unreasonable tuning requests (e.g., requests that would cause a waste of resources and requests for finished queries). The runtime DOP tuning module encompasses a dynamic optimizer and a dynamic scheduler. Upon receiving a tuning request, the auto tuner will generate tuning actions to the dynamic optimizer, which determines the type of DOP tuning required and invokes the dynamic scheduler to perform the tuning operations. \cref{chouxiang} illustrates the two types of DOP tuning available in \sysname: \textbf{intra-task DOP tuning} (\textcircled{1}), which involves changing the number of drivers for a pipeline (detailed in \cref{ITRE}), and \textbf{intra-stage DOP tuning} (\textcircled{2}), which involves changing the number of tasks for a stage (detailed in \cref{ISRE}). In the next section, we describe how we implement these new features.

\section{Intra-Query Runtime Elasticity}
\label{IQRESection}

In this section, we focus on how to address the challenges mentioned in \cref{background} to implement IQRE.  \cref{solution} provides a solution overview of IQRE. \cref{bufferRedesign} describes the redesign of buffers for efficient stage DOP tuning. \cref{ITRE} and \cref{ISRE} show the process of tuning task DOP and stage DOP, respectively. \cref{hashjoinDOPTuning} discusses the parallelism tuning for hash join. \cref{elasticshuffle} describes how to use runtime elasticity to reduce shuffle overhead.

\subsection{Solution of Runtime Elasticity}
\label{solution}

We now analyze the overall solution for runtime elasticity. Operators in query plans can be classified into two types: stateless and stateful. Stateless operators process pages without relying on any state, directly generating output pages from input pages. In contrast, stateful operators depend on external or historical data to compute output and cannot derive results solely from input pages.
In \sysname, stateless operators include filter, project, sink, source, exchange, task output, and table scan. If a stage or pipeline consists of stateless operators only, we can freely adjust its DOP by generating tasks or drivers dynamically.
%tasks or drivers can be dynamically generated at runtime without any risk.

Stateful operators in \sysname include aggregation (aggregation operator) and join (hash join operator and cross join operator). The aggregation operator maintains global data, which limits the flexibility to modify the parallelism of the task or stage it resides in. To enable runtime elasticity, we adopt a two-stage aggregation model  \cite{pgagg,prestoagg}, similar to \presto. This model divides the aggregation into a partial and final aggregation. The partial aggregation operator handles group-by and pre-aggregation operations, and since its state data can be destroyed and reconstructed, it is considered stateless. The final aggregation operator is stateful: it merges all the partial results with its task and stage parallelism fixed at 1. For join operators, probe-side data processing must wait for the build-side to complete before it can begin. In tasks containing join operations, we focus on adjusting the parallelism of the probe pipeline. When the hash table building is finished on the build side, the probe pipeline can freely generate and close drivers. However, increasing the parallelism for the stage containing the join operation requires hash table repartition/reconstruction, which we will discuss in detail in \cref{hashjoinDOPTuning}.

\subsection{Redesign of Buffers}
\label{bufferRedesign}

As previously mentioned, generating new tasks for a stage requires adjusting numerous components of both upstream and downstream stages. To ensure efficiency and robustness in stage DOP adjustments, we confine the scope of components affected by parallelism modifications to the upstream and downstream buffers. We made significant enhancements to the task output buffer, redistributing more responsibilities to it and enabling its capacity to dynamically adjust as the DOP of the downstream stage changes.

\subsubsection{Redesign of Task Output Buffer}
\label{taskoutputbuffer}

The task output buffer is now responsible for data distribution, shuffling, and parallelism variation adaptation, while the task output operator focuses solely on page delivery. This design ensures that when downstream parallelism changes, the task output buffer can quickly detect new downstream tasks and update the data allocation scheme accordingly. It resembles a shuffle service found in big data frameworks, such as the Spark shuffle service \cite{shuffleService} and BigQuery shuffle service \cite{bigQueryShuffleService}. A shuffling service typically consists of a shuffling cluster that receives intermediate data generated by other clusters (e.g., a Spark cluster) to assist in performing shuffling operations. Additionally, shuffle services can perform dynamic optimizations, leveraging technologies like Adaptive Query Execution (AQE) \cite{sparkAdaptiveExecutionWeb} to determine appropriate parallelism for subsequent job execution stages. However, AQE can only adjust parallelism for a stage after the completion of the previous stage and does not allow for DOP modifications during data processing. In contrast, \sysname can alter stage DOP at any moment (but we believe that adaptive query execution is quite orthogonal to IQRE, and they can be applied simultaneously in one system).

\begin{figure}
  \includegraphics[width=1\linewidth]{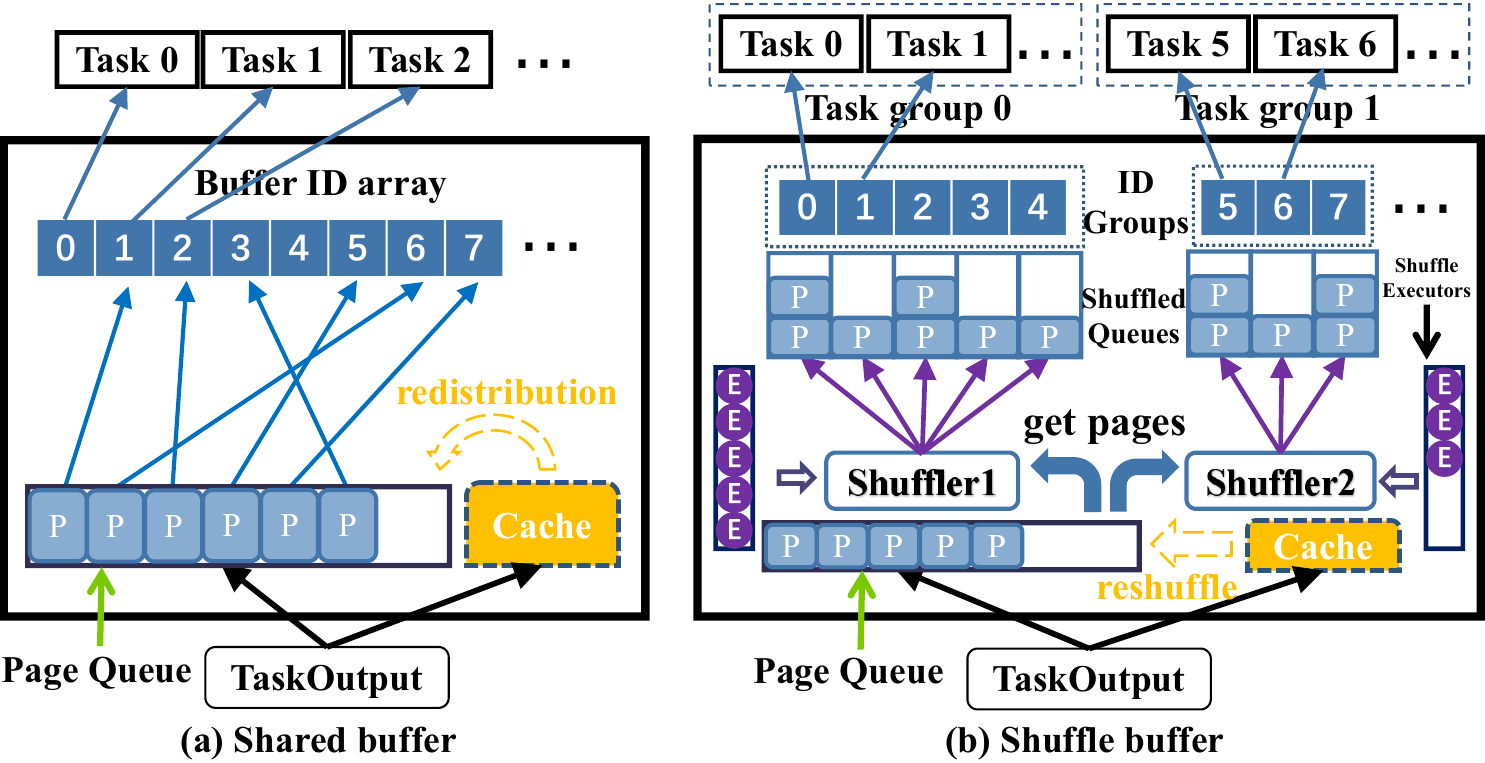}
  \caption{Shared buffer and shuffle buffer. }
  \label{buffers}
\end{figure}

\begin{figure}
  \includegraphics[width=0.4\textwidth]{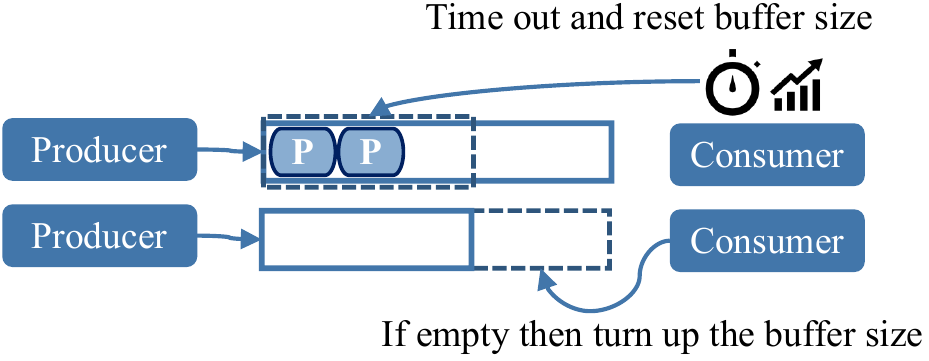}
  \caption{Runtime elastic buffer\textnormal{ -- the consumer side automatically adjusts the buffer capacity to ensure that the rate of data generation matches the rate of data consumption.}}
  \label{buffertune}
\end{figure}

\sysname currently features two types of output buffers: shared buffers and shuffle buffers. As illustrated in \cref{buffers}, both buffers contain a page queue and a page cache. All pages produced by a task are stored in the page queue via the task output operator. The page cache, which is not always necessary, is utilized for reshuffling or redistributing pages for the join build side. The page queue is implemented using TBB's concurrent queue \cite{tbb} to facilitate efficient concurrent access.

Each downstream task retrieves pages using a buffer ID. The Buffer ID array can dynamically change in response to fluctuations in the number of upstream tasks. The shuffle buffer employs shufflers to process pages, with each shuffler containing multiple shuffle executors—threads that perform shuffling operations. The number of executors corresponds to the number of downstream tasks. Each shuffled page queue is linked to a specific buffer ID. And buffer IDs are grouped according to the shuffler to which they belong to form buffer ID groups. And the downstream tasks corresponding to the buffer ID group form task groups.

\subsubsection{Runtime Elastic Buffer}
\label{runtimeelasticbuffer}

As mentioned before, to prevent the buffer capacity from affecting the query execution, we designed the runtime elastic buffer. The buffer capacity is adjusted dynamically by the consumer side at runtime. As illustrated in \cref{buffertune}, if the consumer detects that the buffer is empty, it indicates that the consumption rate is exceeding the production rate. In this case, the consumer will increase the buffer size to accommodate more pages generated or requested by the producer. To align the buffer size with the consumption rate, the consumer periodically (e.g., every 500 milliseconds) counts the number of pages processed and uses this data to resize the buffer. This means that the consumer can determine the optimal amount of data to cache based on its recent consumption capabilities. Since the buffer size is adjusted in real time, we can initially set all buffer capacities to the size of a page.

\subsection{Intra-Task Runtime DOP Tuning}
\label{ITRE}

\begin{figure}
  \centering
  \subfloat[Exchange pipeline and task output pipeline.  \label{intraTask1}]{
    \includegraphics[width=0.41\linewidth]{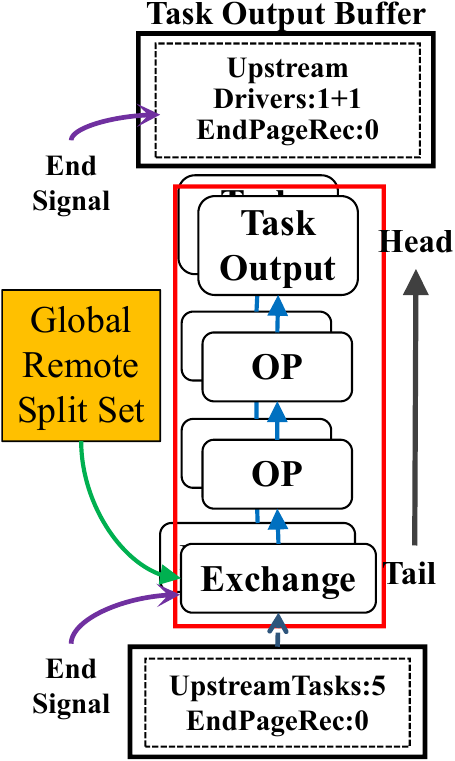}
  }\hspace{0.03\linewidth}
  \subfloat[Sink pipeline and source pipeline. \label{intraTask2}]{
    \includegraphics[width=0.44\linewidth]{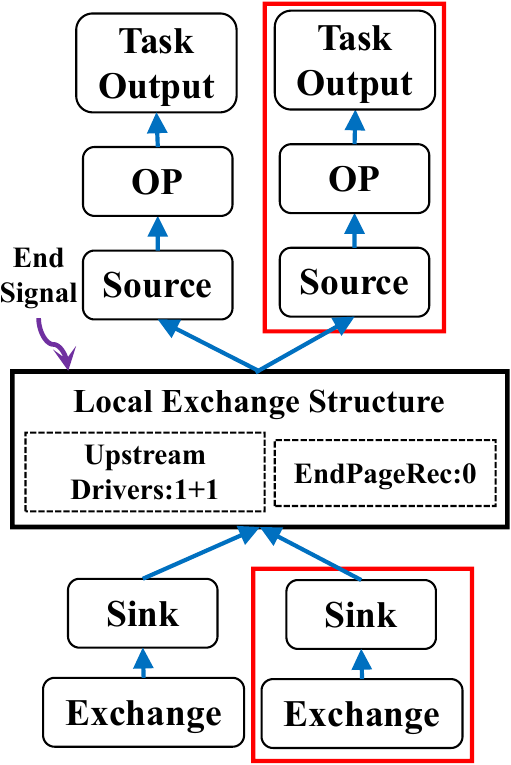}
  }

  \caption{Intra-task DOP tuning \textnormal{ -- the driver in the red box is the newly generated driver.}}
  \label{pipelinetypes}
\end{figure}

\begin{figure}
  \includegraphics[scale=0.45]{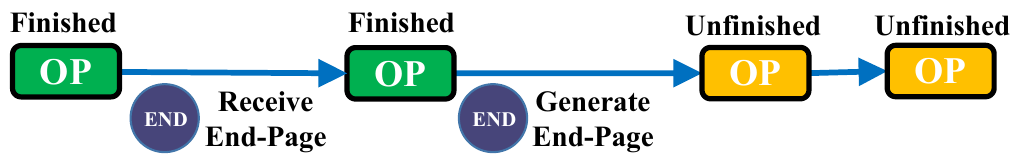}
  \caption{End page relay game \textnormal{--  the end page is passed between operators to gracefully close a driver.}}

  \label{eprelay}
\end{figure}

This section describes how to adjust the task parallelism. Tuning intra-task DOP involves adjusting the number of drivers for a pipeline within a task.

\textbf{Increasing task DOP.} Adding task DOP involves generating new drivers for pipelines, but certain data must be preserved to ensure logical correctness during tuning. As shown in \cref{intraTask1}, for the exchange pipeline (pipeline containing exchange operator), the task maintains a global remote split set which saves all the remote splits the current task uses. When a new driver is created, these splits are directly assigned to the new exchange operator, bypassing the need for coordinator involvement. For the task output pipelines (pipeline containing task output operator) and source pipelines (pipeline containing source operator), it is necessary to track the number of upstream pipeline drivers by recording the number of head physical operators in the upstream pipeline. This record helps determine if the upstream pipeline has completed processing.

\textbf{End page.} In \presto, the end page is primarily used to ensure that downstream stages conclude normally after data processing is complete. \sysname extends this functionality by using the end page to safely shut down one or more tasks or drivers during data processing. The end page can be generated by the table scan operator, the task output buffer, the exchange operator, or the local exchange structure. By sending ``end signals'' to these components, we effectively manage the shutdown of drivers and tasks. As depicted in \cref{eprelay}, when an operator within the driver receives the end page, a stateless operator will enter the finished state and pass the end page to the next operator. In contrast, a stateful operator must wait until all results are output before entering the finished state and passing the end page along. The end page is transmitted between operators, facilitating normal driver shutdowns.

\textbf{Decreasing task DOP.} We utilize an end signal to shut down drivers. For the exchange pipeline, upon receiving the end signal, the exchange operator halts data reception and adds an end page to the exchange buffer. If we want to decrease the source pipeline parallelism, we let the task send an end signal to the corresponding local exchange structure, which then generates end pages and relays them to source operators. If any component—whether the exchange operator, local exchange structure, or task output buffer—detects that upstream execution is complete (i.e., the number of received end pages matches the number of upstream drivers), it broadcasts end pages to the downstream components.

\subsection{Intra-Stage Runtime DOP Tuning}
\label{ISRE}

\begin{figure}
  \centering
 
    \includegraphics[width=0.7\linewidth]{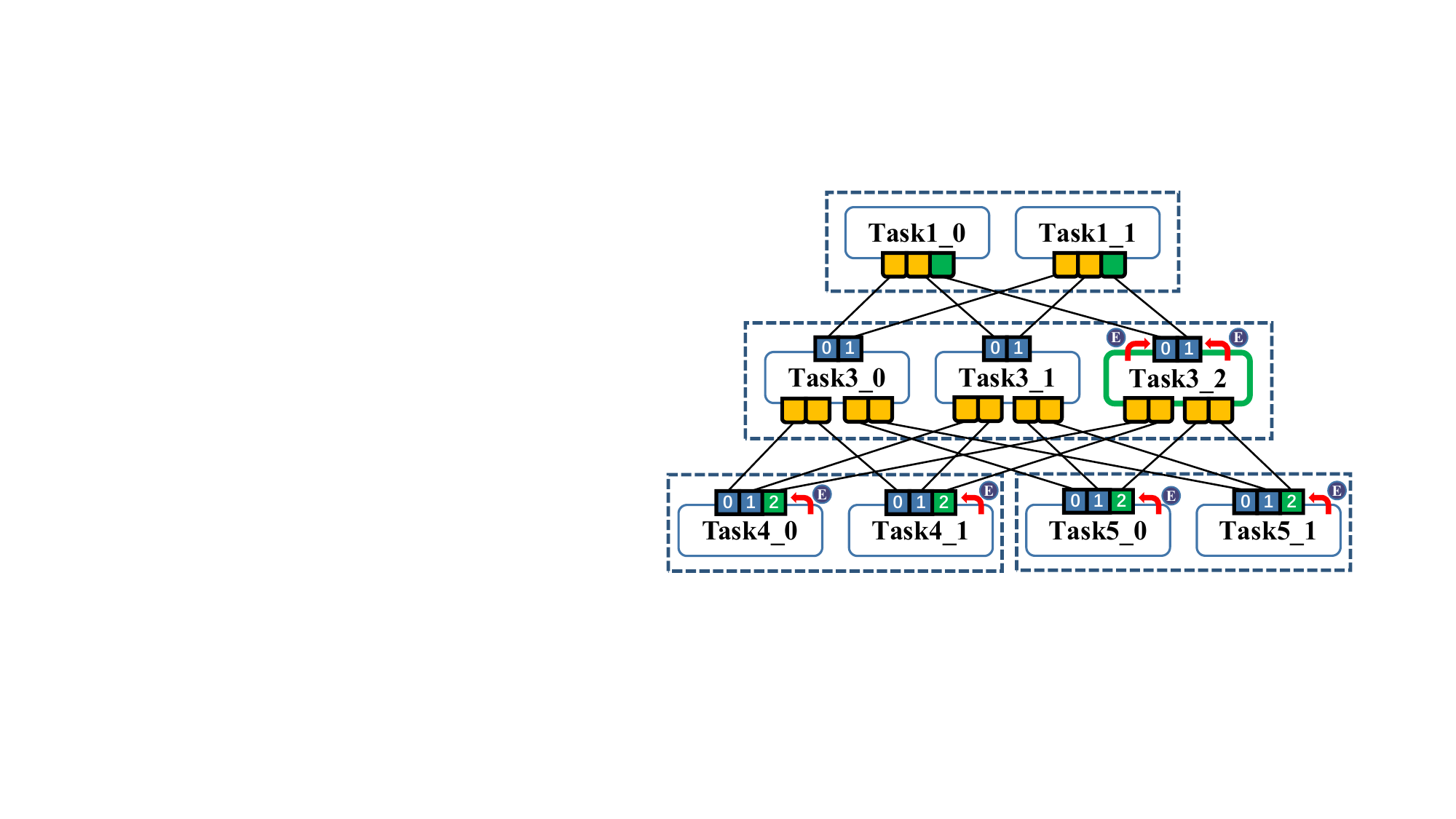}

\caption{Stage DOP tuning \textnormal{-- increasing or decreasing the parallelism for stage 3 in \cref{exexecutionplan}.}}
\label{addConcurrentSimpleBuffers}
\end{figure}

This section describes how to adjust the stage parallelism. Intra-stage DOP runtime tuning involves adjusting the number of tasks within a stage.

During the query scheduling phase, the scheduler constructs an initial distributed execution plan based on the stage tree, traversing it in a bottom-up manner to generate tasks for each stage and establish communication links between them. The dynamic scheduler then tunes the DOP for each stage within this execution plan. \cref{addConcurrentSimpleBuffers} presents the stage DOP tuning process on the partial execution plan for the query depicted in \cref{exexecutionplan}. Below, we detail the process of adding tasks to an execution plan.

\textbf{Increasing stage DOP.} Enhancing the stage DOP involves three steps: 1. Generating a new task (task3\_2) for the stage (stage 3). 2. Provide the address of the new task (including the worker node's IP and task ID) to the parent stage tasks (task1\_0 and task1\_1). 3. Setting the addresses of the child stage tasks (task4\_0, task4\_1, task5\_0, and task5\_1) for the new task.

\textbf{Decreasing stage DOP.} As mentioned before, the end page is used to close tasks. As shown in the \cref{addConcurrentSimpleBuffers}. If we want to close task3\_2, the dynamic scheduler sends end signals to the task output buffers (buffer ID 2) of stage 3's child stages. End pages are generated and passed through task3\_2 to task1\_0 and task1\_1. Task1\_0 and task1\_1 delete the RPC address of task3\_2 and then task3\_2 is destroyed.

\begin{figure}
  \centering
 
    \includegraphics[width=0.7\linewidth]{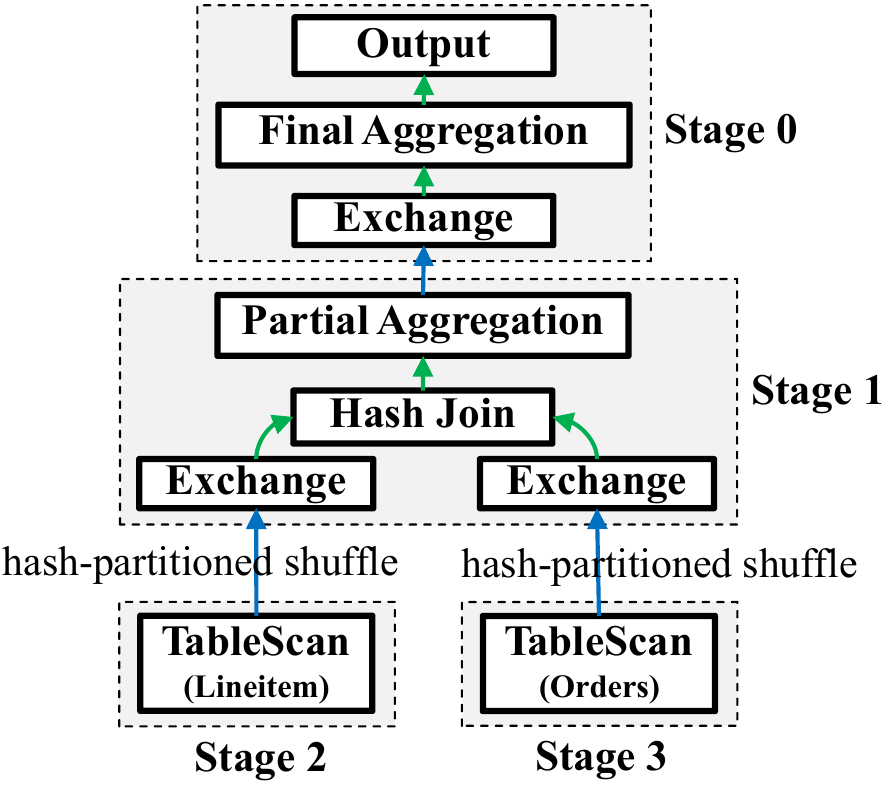}
 
  \caption{The distributed physical plan of the two-way join.}
  \label{2joinplan}
\end{figure}

\begin{figure}
  \centering
  \subfloat[Broadcast join. \label{broadcastjoin}]{
    \includegraphics[width=0.328\linewidth]{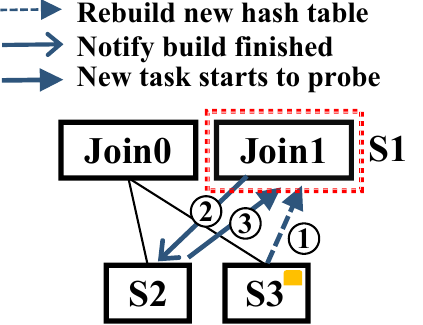}
  }\hspace{0.02\linewidth}
  \subfloat[Partitioned hash join. \label{hashpartitionjoin}]{
    \includegraphics[width=0.605\linewidth]{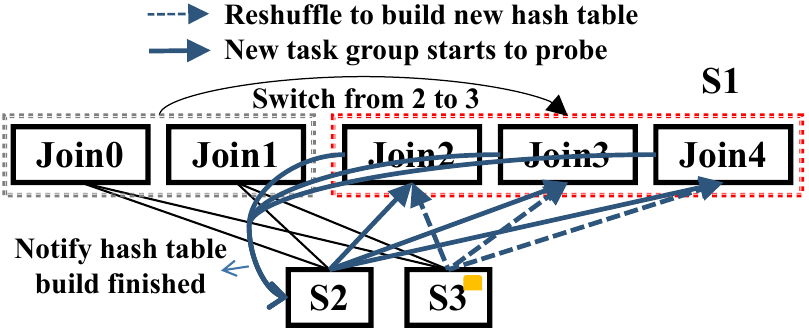}
  }

  \caption{DOP tuning of broadcast join stage and partitioned hash join stage.}
  \label{hashjoinDOPtuning}
\end{figure}

\subsection{DOP Switching for Partitioned Hash Join}
\label{hashjoinDOPTuning}

This section analyzes the runtime elasticity of hash joins, which can be categorized into two types: broadcast hash join and partitioned hash join.

Changing the DOP of a stage containing a join operation requires hash table reconstruction. For partitioned hash joins, the hash table is distributed across multiple tasks, complicating parallelism tuning.

Consider a two-way join query:
\begin{lstlisting}[language=SQL]
SELECT count(l_orderkey) FROM Lineitem INNER
JOIN Orders ON Lineitem.orderkey = Orders.orderkey
\end{lstlisting}

\cref{2joinplan} illustrates the distributed physical plan of the two-way join. \cref{hashjoinDOPtuning} presents two partial execution plans (left for broadcast join and right for partitioned hash join) for the two-way join. Each rectangle represents a task. As shown in \cref{broadcastjoin}, increasing the parallelism of stage 1 simply involves generating a new task (Join1) and reconstructing a new hash table on Join1 via stage 3. For partitioned hash join, we implement parallelism modifications using a method called ``DOP switching''. This entails the build side (stage 3) first creating a new distributed hash table in a new task group, after which the probe side utilizes this new task group for the remaining join operations (the previous task group is closed).

A critical challenge is efficiently building a new distributed hash table. An intuitive solution is to re-balance the distributed hash table from the previous task group, a method employed in various works \cite{snowflake-rebalance, dynahash}. However, we argue that this approach is unsuitable for query DOP tuning, as re-balancing can disrupt probe operations, leading to increased query latency. Instead, ``rebuilding the hash table by the upstream stage'' is more robust and minimizes disruption to query execution. To optimize DOP tuning, we ensure that the probe side only switches DOP after the new task group completes the hash table construction.

\begin{figure}[t]
  \centering
 
    \includegraphics[width=0.9\linewidth]{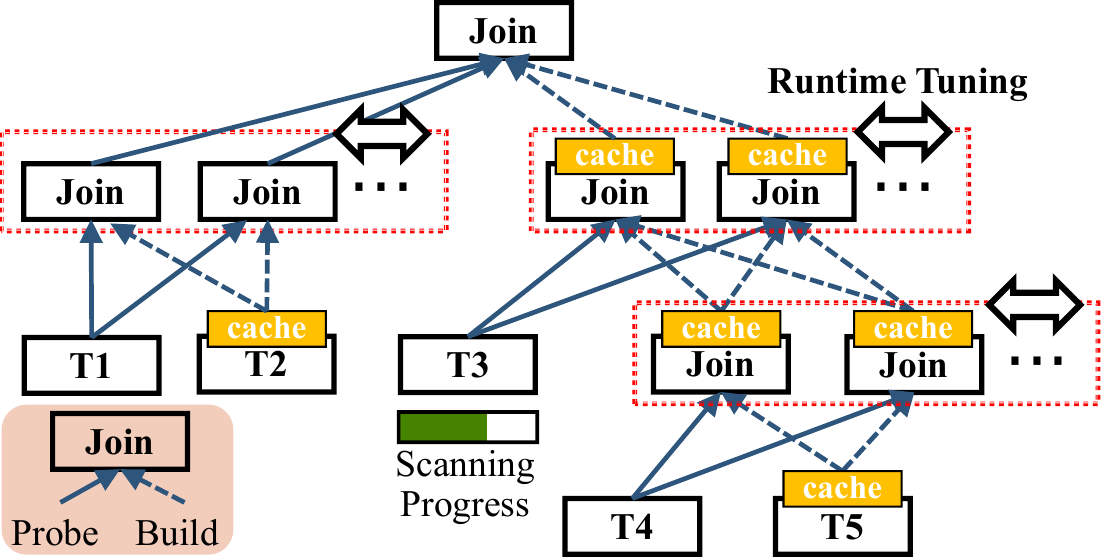}

  \caption{Intermediate data caching \textnormal{-- hash join relies on intermediate data caches to implement parallelism tuning. The remaining execution time of a stage can be predicted by the execution progress of the table scan stage it depends on.}}

  \label{joinoverview}
\end{figure}

We employ intermediate data caching to facilitate this method for multi-table joins. Specifically, the build-side stage temporarily stores intermediate results for subsequent reuse. As depicted in \cref{joinoverview}, the new distributed hash table can be created from the intermediate data cache of the upstream stage. This caching technique is widely utilized in distributed systems (e.g., Snowflake, Redshift) to significantly reduce query latency. In \presto, this mechanism is referred to as fragment result caching \cite{resultCaching}.

\subsection{Elastic Shuffle Stage}
\label{elasticshuffle}

The shuffle operation can easily become a bottleneck for partitioned hash joins, and reshuffling can significantly impact the efficiency of DOP switching. The solution is to increase the number of nodes involved in the shuffling. There are two primary methods to reduce shuffle latency: 1. Distributing data across more compute/storage nodes. 2. Inserting a shuffle stage downstream of the table scan stage. Users can adjust the shuffle rate by tuning the parallelism of the shuffle stage at runtime. The shuffle stage consists solely of a pipeline comprising an exchange operator and a task output operator, with the shuffle buffer performing the shuffle operations.

\section{Automatic DOP Tuning}
\label{strategies}

\sysname incorporates an auto-tuner designed to optimize the DOP of a query automatically without users' attention. It also provides a user-friendly interface for tuning the query DOP manually to understand the effect of each parallelism adjustment. Users can interact with the auto-tuner via buttons, which guide them in adjusting parallelism effectively with what-if service.

The implementation of the auto-tuner relies on three components: runtime bottleneck localization, stage remaining execution time prediction, and DOP tuning request filter. Runtime bottleneck localization means the system identifies stage IDs that require adjustment based on the execution progress of the query—these stages are computational bottlenecks. Additionally, if the query encounters non-computational bottlenecks (e.g., network bottleneck), the system can detect and highlight these as well. The stage remaining execution time prediction informs users of the expected remaining execution time for a stage when parallelism is modified, facilitating better user decision-making based on their needs. The DOP tuning request filter is used to filter invalid or inefficient parallelism tuning requests.

\subsection{Runtime Bottleneck Localization}

We identify computational bottlenecks by adding special counters to the exchange buffer. Suppose a stage is not a computational bottleneck. In that case, it indicates that the page processing rate of tasks in this stage exceeds the page producing rate of the upstream stage, resulting in this stage's tasks' exchange buffers often being empty. Conversely, a computational bottleneck stage will typically have a populated exchange buffer. As discussed in \cref{runtimeelasticbuffer}, when the exchange buffer becomes empty, the consumer side increases the buffer size. We let each task maintain a turn-up counter. For each increase, the turn-up counter increments by one. So, If the counter's value remains unchanged during stage execution, we classify this stage as a computational bottleneck.

\begin{figure}
  \centering
 
    \includegraphics[width=0.9\linewidth]{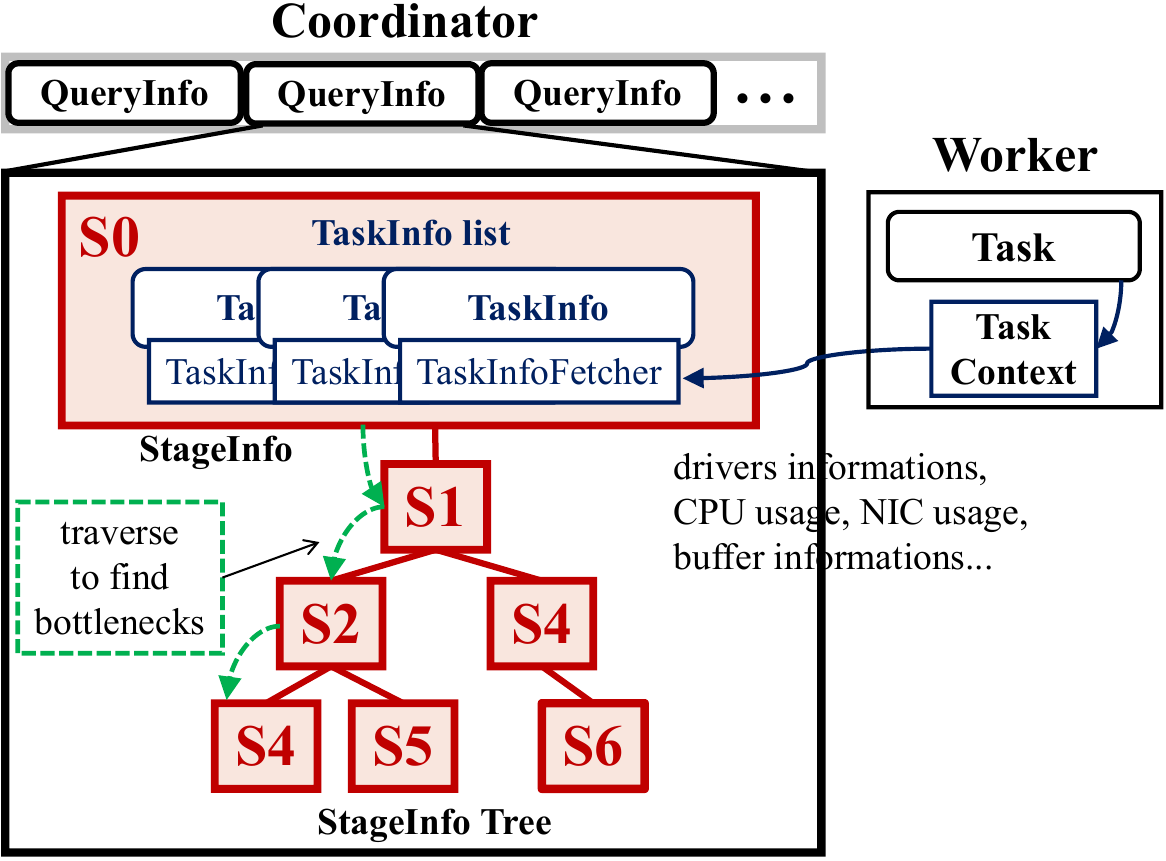}
 
  \caption{Query runtime information collection \textnormal{-- runtime information of queries is organized to multiple levels. The coordinator finds the query bottleneck by traversing the stage tree.}}
  \label{infocollector}
\end{figure}

\sysname collects and organizes query runtime information using a "query-stage-task" hierarchical structure, as illustrated in \cref{infocollector}. Each task stores its own runtime information in the task context, and the coordinator's runtime information collector periodically collects this information by task information fetchers from tasks' contexts. This information is grouped and aggregated by stage and query to support decision-making. This contains the counter information mentioned above. When the coordinator receives the user's prediction request, it goes through the entire stage info tree and locates the stage bottleneck based on the information recorded. 
The coordinator also monitors other metrics, such as the NIC utilization to determine if a stage is experiencing a network bottleneck.
%The coordinator also monitors other metrics, such as the NIC utilization of each worker node. This information helps the coordinator determine if a stage is experiencing a network bottleneck by comparing the metric with maximum NIC bandwidth.

\subsection{DOP Tuning Request Filter}
\label{remainingtime}

In some scenarios, tuning parallelism may be ineffective. The DOP tuning filter is designed to block such inappropriate requests. Currently, the filter handles two types of requests: 1. parallelism adjustment requests for queries or stages that have already been finished, 2. unsuitable requests for stages containing join operations. For example, if a stage is close to completion and the time required to rebuild the hash table exceeds the remaining execution time, adjusting the parallelism would be a waste of resources.

To realize this, we need to estimate the remaining execution time for a stage and compare it with the hash table construction time. We illustrate this with the example in \cref{joinoverview}. Since a stage has multiple tasks (each task has a hash table build time), we represent the hash table build time for the stage by the maximum hash table build time of its tasks. To predict the remaining time, we monitor the progress \cite{progress1, progress2} of the stage’s execution. In this paper, we leverage the table scanning progress of the table scan stage (upstream stage of probe side) to predict the remaining execution time for the join stage. The coordinator periodically records the remaining data volume ($V_{remain}$) of the table scan stage and calculates the data consumption rate ($R_{consume}$). The remaining time can then be estimated as $ T_{remain} = V_{remain}/R_{consume}$. If the estimated remaining time is less than the hash table construction time, the DOP tuning request is rejected.

Below we explain why it is sufficient to compute only the progress of the table scan stage. In fact, the query progress on the \sysname main UI only shows the progress of each table scan stage. Given that query execution processes data in a streaming fashion, data from the table scan stage is incrementally passed to downstream stages rather than all at once. Each intermediate stage retrieves a limited number of pages from the table scan stage at a rate aligned with its own processing capacity, thereby avoiding the problem of excessive data caching. Consequently, the rate at which data is consumed in the table scan stage serves as a reliable approximation of overall query execution progress.

\subsection{Stage Remaining Execution Time Prediction}

We employ a straightforward principle to predict the remaining execution time of a stage. Specifically, if the DOP of a target stage is scaled up by a factor of $n$, then the throughput of its upstream stage must also scale up by the same factor. In \cref{remainingtime}, we outlined how to calculate the remaining execution time $T_{remain}$ of a stage. Assume that the current parallelism of the target stage is $n_1$, and the desired parallelism is $n_2$, where $n_2 > n_1$, the factor for the increase in parallelism is $n_f = n_2/n_1$. If the throughput of the current stage can indeed increase by a factor of $n_f$, we predict the remaining execution time of the current stage as follows: $T_{predicted} = (T_{remain}-T_{tuning})/n_f$. Here, $T_{tuning}$ refers to the time needed for parallelism adjustment. If the stage does not involve join operators, then $T_{tuning} \approx  0$. However, if the stage includes join operators, $T_{tuning} \approx T_{build}$, where $T_{build}$ represents the time required for hash table reconstruction.

However, $n_f$ cannot be arbitrarily large values in practice. 
%The throughput of the upstream stage generally cannot scale up by the same factor. 
The maximum $n_f$ is influenced by the upstream stage's CPU and network utilization, among other factors. If the upstream throughput rate is affected by CPU utilization, we can use the remaining CPU resources and the current CPU utilization of the upstream stage to estimate a maximum $n_f$. This value is calculated in real-time from data collected by the runtime information collector. Estimating $n_f$ helps prevent unreasonable parallelism adjustments, such as increasing stage parallelism by a factor of 1000. When a user requests an increase in parallelism by a factor of $n$, the coordinator first calculates $n_f$ based on runtime data. If $n < n_f$, the coordinator uses $n$ to compute the remaining time; otherwise, it uses $n_f$ directly for the calculation.

\begin{figure}
  \includegraphics[width=0.9\linewidth]{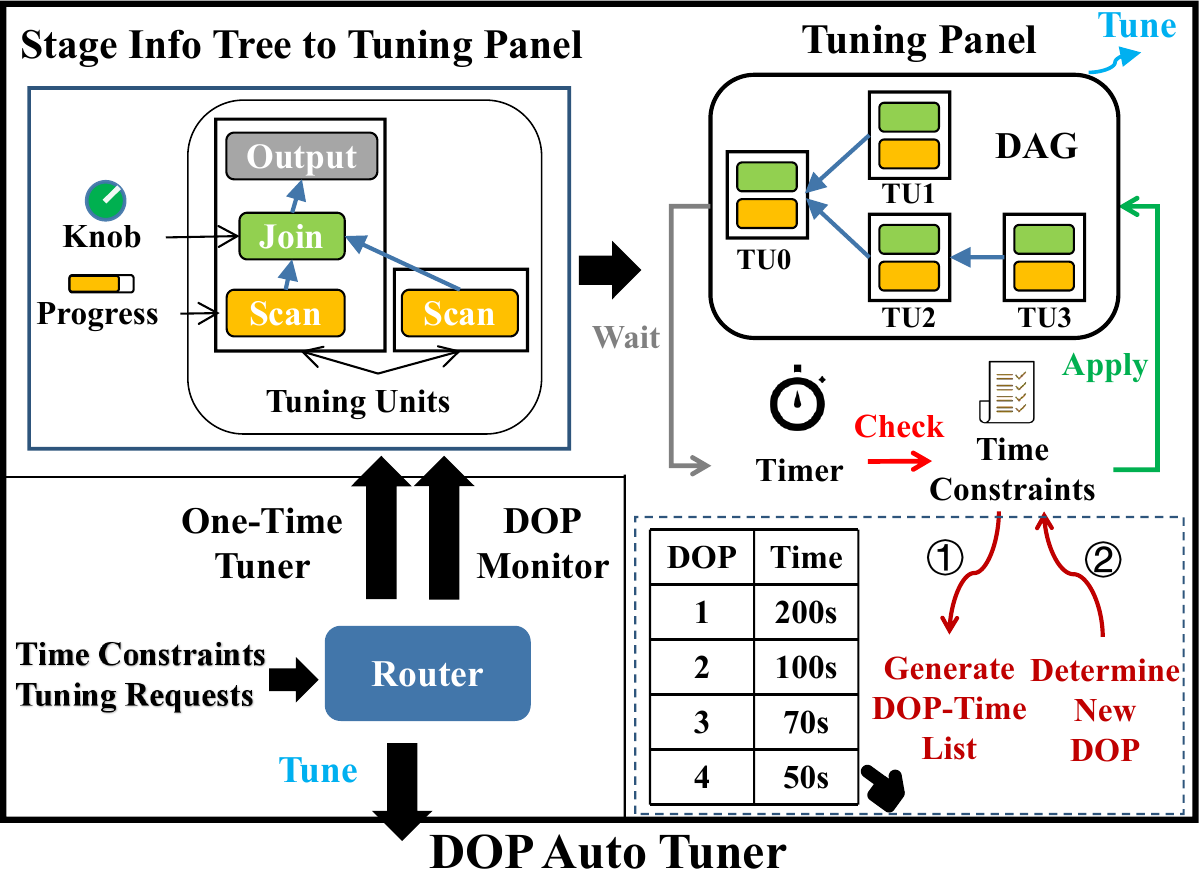}
  \caption{Automatic DOP Tuning Workflow.}
  \label{autoTuner}
\end{figure}

\subsection{DOP Auto-Tuner}

In this section, we describe the auto-tuner in detail (\cref{autoTuner}). The DOP auto-tuner supports three types of requests: direct DOP tuning (manual adjustment of DOP), one-time auto-tuning (tuning the stage DOP once based on latency constraint),  and DOP monitor (periodically checking stage execution progress to adjust DOP).

The auto-tuner decomposes the query stage info tree into multiple DOP tuning units. Each unit comprises a progress indicator (at the table scanning stage) and tuning knobs (intermediate stages with adjustable parallelism). These units collectively form an execution Directed Acyclic Graph (DAG), presented as a DOP tuning panel. By leveraging the DAG, the auto-tuner monitors query execution progress and dynamically operates the tuning knobs according to the time constraints.

Upon receiving a tuning request, the auto-tuner predicts the remaining execution time for the target stage and generates a DOP-time list that estimates the stage's execution time at various DOP configurations. It then selects the DOP configuration that most closely aligns with the query latency constraint and applies the adjustment via the tuning panel. Users can also enable the DOP monitor (\cref{autoTuner}), especially for long-running queries, that will periodically track the execution progress of each stage and incrementally adjust the DOP to meet the query's latency constraint while minimizing resource usage.

% \revision{} {Upon receiving a tuning request, the auto-tuner predicts the remaining execution time for the target stage and generates a DOP-time list that estimates the stage's execution time at various DOP configurations. It then selects the DOP configuration that most closely aligns with the query latency constraint and applies the adjustment via the tuning panel (increase parallelism if time is tight, decrease parallelism if time is loose). For stages with prolonged execution times (e.g., exceeding ten minutes), throughput fluctuations may cause actual execution times to diverge from predictions. To address this, users can enable a DOP monitor (\cref{autoTuner}), which periodically tracks the execution progress of each stage. The monitor incrementally adjusts the DOP to align query execution within time constraints, ensuring timely completion while minimizing resource usage.}

\begin{figure}
  \includegraphics[width=0.9\linewidth]{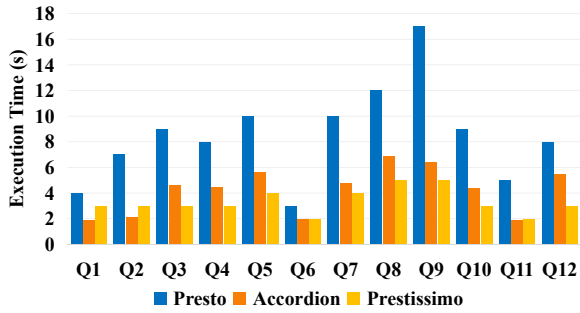}
  \caption{Standalone TPC-H benchmark results \textnormal{-- for \sysname, \presto , and \prestissimo with scale factor of 1.}}
  \label{benchmark}
\end{figure}

\section{Evaluation}
\label{evaluation}

In this section, we evaluate the efficiency of IQRE of \sysname. \cref{experiment} details the experimental setup. \cref{ITRS_Exper} evaluates the intra-task runtime elasticity, while \cref{ISRS_Exper} evaluates the stage runtime elasticity. \cref{hash_Exper} evaluates the DOP switching and elastic shuffle stage performance. Finally, \cref{prediction}  demonstrates the effect of DOP auto-tuning.

\subsection{Experimental Setup}
\label{experiment}

We conducted experiments on a cluster of 21 AWS EC2 (c5.2xlarge) nodes, each node equipped with 16GB of RAM, and 30GB SSD, with a 10Gbps NIC bandwidth. The cluster comprises 1 coordinator node, 10 storage nodes, and 10 compute nodes.

We first tested \sysname's benchmark (as shown in \cref{benchmark}) with 12 TPC-H queries (SF1) on a single node and compared it to  \presto and  \prestissimo (the C++ version of \presto) to verify that the system implementation is reasonable. Then we performed experiments on TPC-H with a scale factor of 100 (\textbf{TPC-H SF100}).

 %\cref{benchmark} presents benchmark results (stand-alone) comparing \sysname with \presto and \prestissimo (the C++ variant of \presto) on the TPC-H benchmark suite ($Q1 \sim Q12$) at a scale factor of 1.

In \presto, the table scan operator can fetch splits from remote sources (such as Hive, AWS S3, etc.) for processing. To eliminate the variability introduced by different data sources and formats, we used CSV format for data storage. The table scan operator reads CSV files via the \arrow CSV file reader (\arrow supports various file formats, including CSV, Parquet, ORC, and so on). Since no remote data source is used in this experiment, the TPC-H tables need to be manually divided into multiple splits before query processing. \cref{tableSplits} outlines the partitioning scheme for each TPC-H table. \sysname includes a built-in scripting language for controlling query initiation and parallelism adjustments at specified times. We use script executor to track throughput variations, manage both parallelism changes and result recording in experiments.

%\sysname features a built-in scripting language that allows control over query initiation and parallelism adjustments at specified time points. For example, the script statement ``BEGIN; START\_QUERY Q3; WAIT\_QUERY 1000; ADD\_PARALLELISM STAGE 3,1; END;'' instructs the script executor to start query Q3, wait for one second, and then increase the parallelism of stage 3 by one. The script executor also records throughput variations throughout the query execution. In this paper's experiments, both the parallelism adjustments and the recording of experimental results are managed by the script executor.

%We employ two queries to demonstrate intra-query runtime elasticity: TPC-H Q3 and a two-way join query Q2J (\cref{2joinplan}). The join type for Q3 is a broadcast join, while Q2J utilizes a partitioned hash join.

\begin{table}
  \caption{TPCH-SF100 Table Setup---Total 107GB}
  \label{tableSplits}
  \begin{tabular}{cccc}
    \toprule
    Table &Partitioning scheme &Table size & Split size\\
    \midrule
    \texttt{Nation} & 1 node, 1 split/node &2.5KB & 2.5KB\\
    \texttt{Region}&1 node, 1 split/node &512B& 512B\\
 \texttt{Supplier} & 10 nodes, 1 split/node &137MB& 13.7MB\\
 \texttt{Part} & 10 nodes, 1 split/node &2.29GB& 0.23GB\\
 \texttt{Partsupp} & 10 nodes, 1 split/node &11.37GB& 1.13GB\\
 \texttt{Customer} & 10 nodes, 1 split/node &2.29GB& 0.23GB\\
 \texttt{Orders} & 10 nodes, 1 split/node &16.57GB& 1.66GB\\
    \texttt{Lineitem}& 10 nodes, 7 splits/node &74GB& 1.06GB\\
    \bottomrule
  \end{tabular}
\end{table}

\begin{figure}
  \centering
 
    \includegraphics[width=0.8\linewidth]{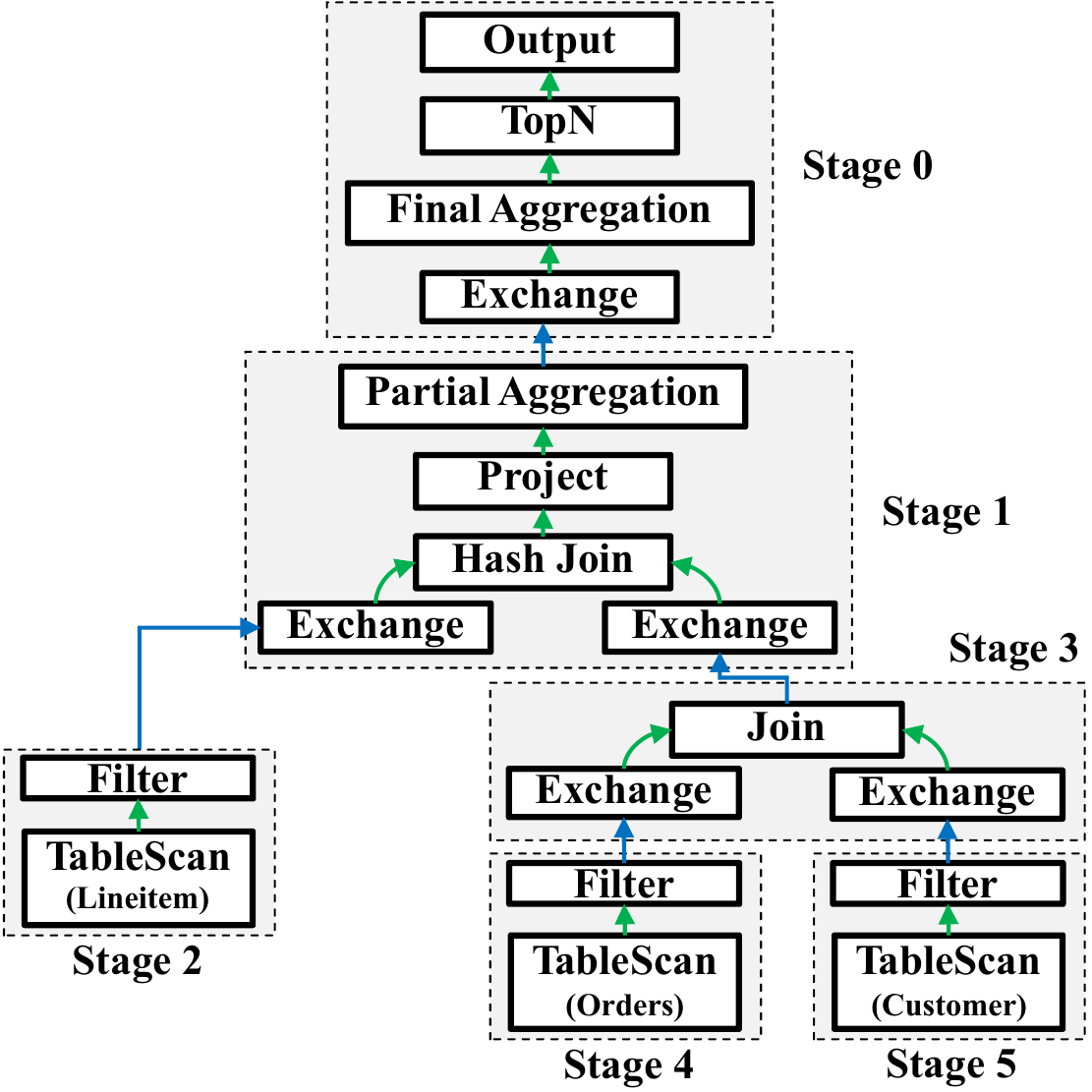}
 
  \caption{The distributed physical plan of the Q3.}
  \label{Q3plan}
\end{figure}

\begin{figure}
  \centering
 \includegraphics[width=0.8\linewidth]{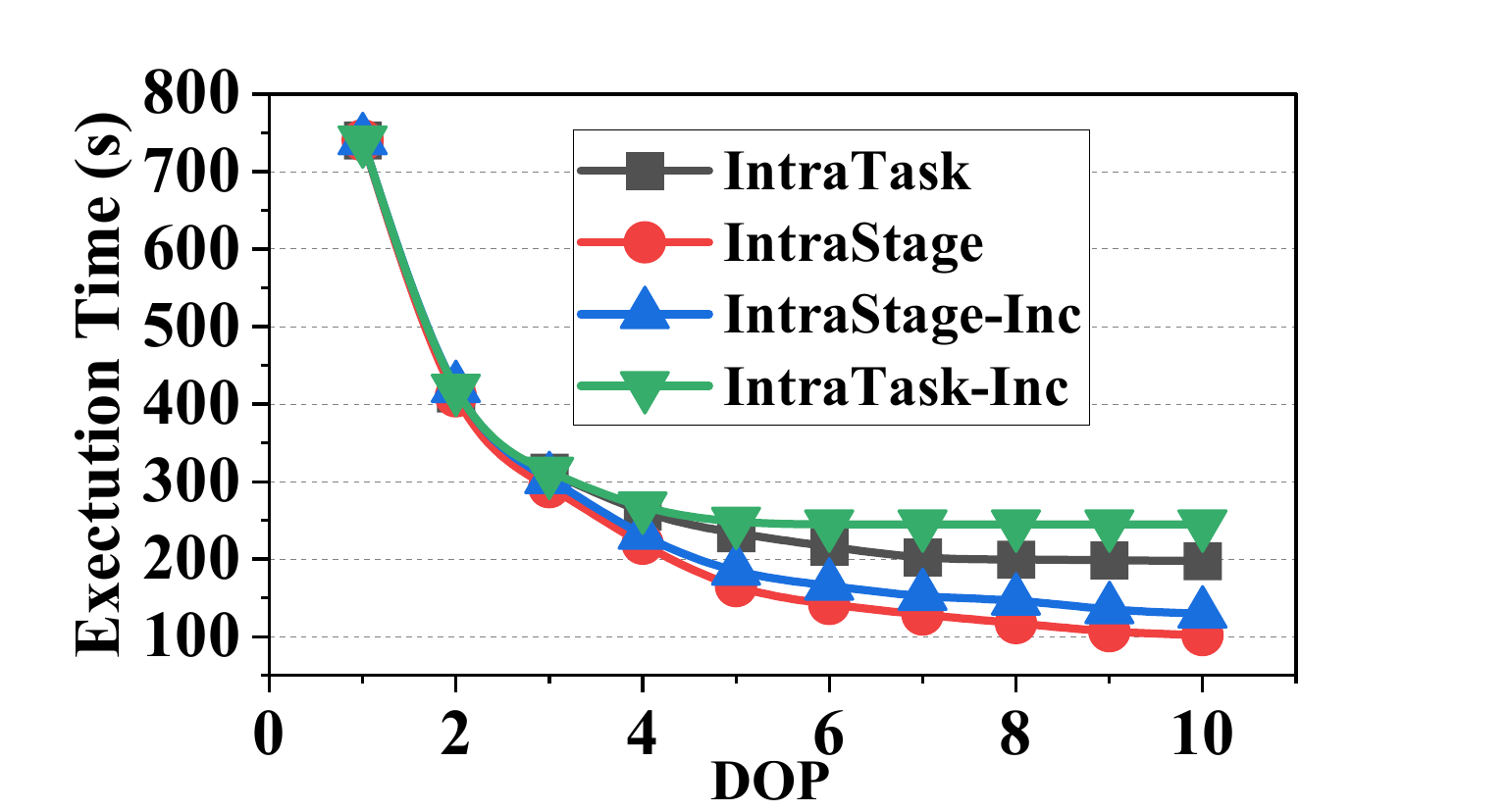}
  
  \caption{The Q3 execution time curves \textnormal{-- with different degrees of intra-stage parallelism and intra-task parallelism.}}
  \label{Q3ExeOrigin}
\end{figure}

\begin{figure}
  \centering

 \includegraphics[width=0.85\linewidth]{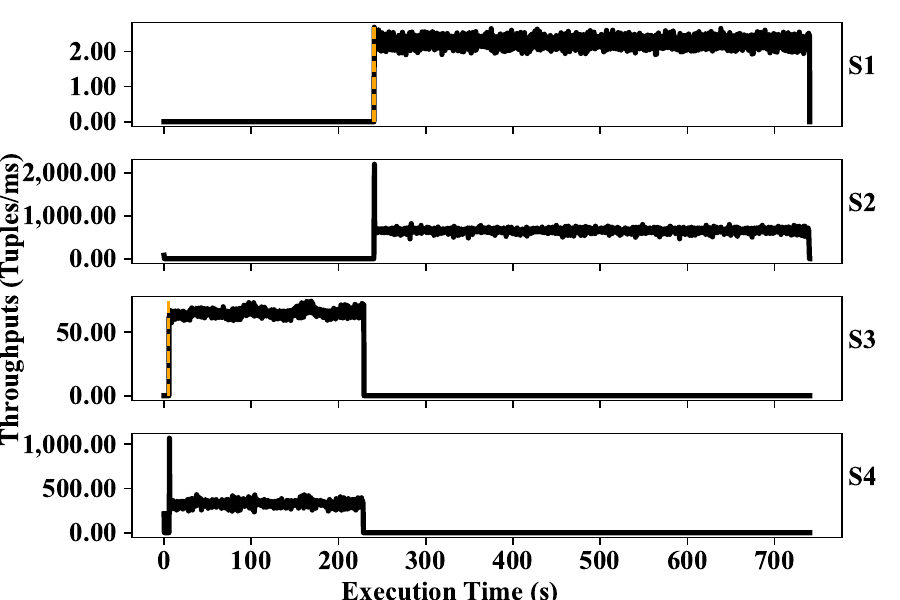}
  
  \caption {The Q3's raw stage throughput curves \textnormal{-- with each stage parallelism of 1.}}
  \label{Q3TPOrigin}
\end{figure}

\subsection{Task DOP Runtime Tuning}
\label{ITRS_Exper}

This section evaluates the intra-task parallelism adjustment. We take TPC-H Q3 as an example to show the evaluation results. \cref{Q3plan} presents the distributed physical plan of Q3, while \cref{Q3ExeOrigin} displays the execution times for Q3 across various degrees of intra-stage and intra-task parallelism (representing \presto-like execution times without runtime adjustments). \cref{Q3TPOrigin} illustrates the throughput variations for each stage of Q3 with stage parallelism of 1, omitting stages 0 and 5 due to their negligible throughput and brief duration.

From \cref{Q3plan}, we observe two types of dependencies between stages: execution dependency, where one stage must be completed before another can start, and data dependency, where a stage requires data from an upstream stage for processing. For instance, stage 2 has a data dependency on stage 1, while stage 3 exhibits an execution dependency on stage 1.

\begin{figure}
  \centering

 \includegraphics[width=0.9\linewidth]{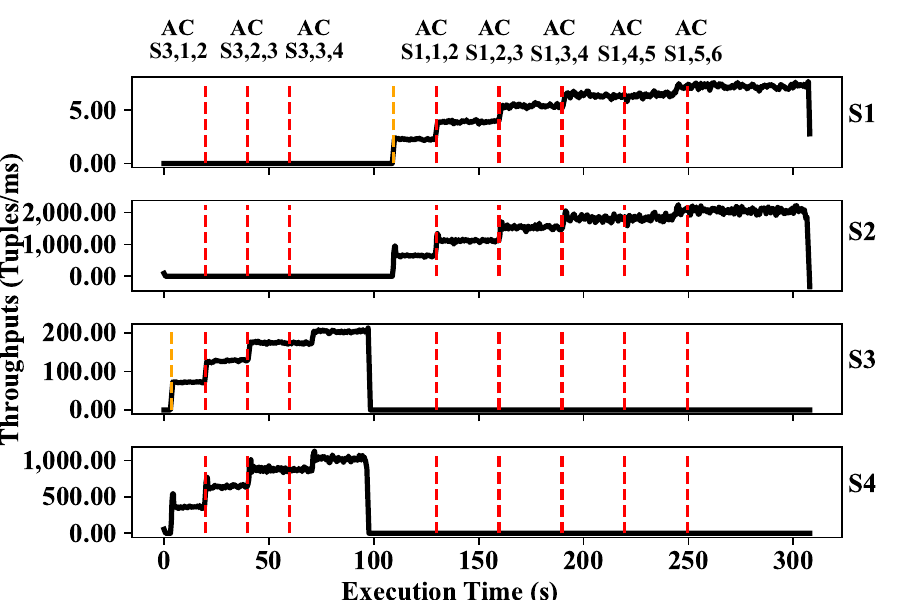}
  
  \caption{The stage throughput curves of intra-task DOP tuning of Q3.}
  \label{Q3IntraTaskTuning}
\end{figure}

\cref{Q3IntraTaskTuning} presents the throughput variations resulting from intra-task DOP tuning for Q3. The initial stage and task parallelism for Q3 are both set to 1. The notation ``AC S$n$, $a$,$b$'' indicates that adding task DOP for all tasks of stage $n$ from $a$ to $b$ at the time marked by the red line. For stages with join operations, yellow dashed lines indicate the completion of hash table construction. The script executor adjusted the DOP for stage 3 twice and for stage 1 three times, progressively increasing throughput with each adjustment. Throughput improves immediately (within 1\~10ms) post-DOP tuning due to rapid physical pipeline generation. Notably, the third adjustment for stage 1 does not enhance throughput, as the first two adjustments already maximized CPU utilization. The total execution time for the query is 307.87 seconds, reflecting a 58.42\% reduction compared to the original execution time of 740.34 seconds (shown in \cref{Q3ExeOrigin}).

%To evaluate the task DOP adjustment overhead, we let Q3 start its execution with the task DOP of 1, and then subsequently increase the parallelism to $n$ and record its final execution time. \cref{Q3ExeOrigin} shows this result (IntraTask-Inc curve). The task DOP tuning overhead mainly consists of the scheduling overhead and the overhead of generating tasks and drivers. We find that for all queries, the overhead of task and driver generation is less than 1 ms. The initial query plan construction for Q3 consumes 65 restful requests, which cost 313ms (In \sysname, each schedule request costs 1\~10ms). So task DOP tuning can immediately change the speed of query execution. The interval between the IntraTask-Inc curve and the Intra-Task curves is caused by scheduling delays.

To assess the overhead associated with task DOP adjustments, we initiated the execution of query Q3 with a task DOP of 1, progressively increasing the parallelism to $n$ while recording its final execution time. The results are depicted in \cref{Q3ExeOrigin} (IntraTask-Inc curve). The overhead of task DOP tuning primarily comprises scheduling overhead and the overhead of generating tasks and drivers. Our analysis reveals that for all queries, the task and driver generation overhead is minimal, consistently below 1 ms. The initial query plan construction for Q3 involves 65 RESTful requests, incurring a total cost of 313 ms (each RESTful request in \sysname takes between 1 and 10 ms). This shows that task DOP tuning can promptly adjust the query execution speed. The observed gap between the IntraTask-Inc curve and the Intra-Task curve is attributable to scheduling delays.

\begin{figure*}
  \centering
  \subfloat[Q3---Initial: 313ms; State transfer: \{S1: 14.11s; S3: 2.99s\} \label{IntraStagetuning}]{
    \includegraphics[width=0.45\linewidth]{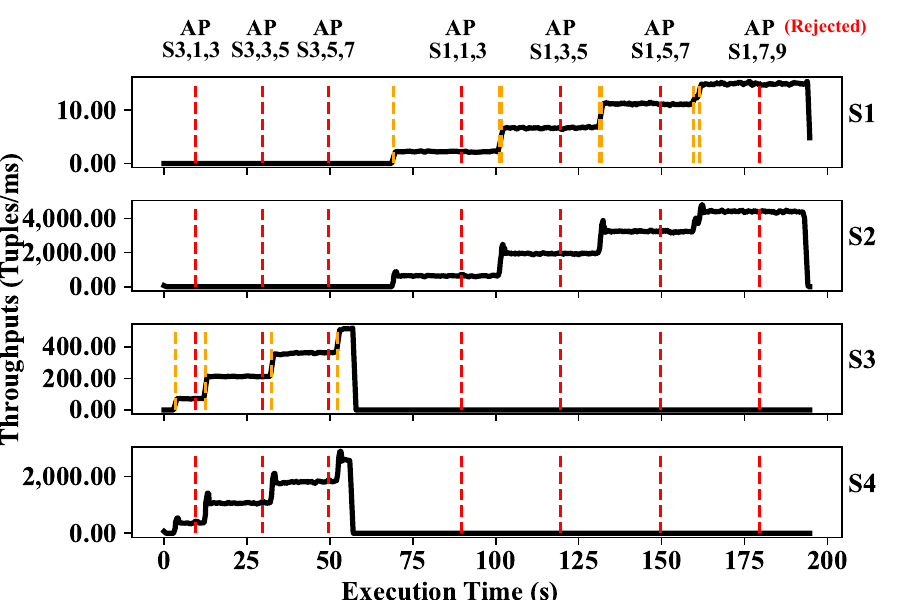}
  }
  \subfloat[Q1---Initial: 156ms; State transfer: \{S1: 6ms\} \label{ISRS-Q1}]{
    \includegraphics[width=0.45\linewidth]{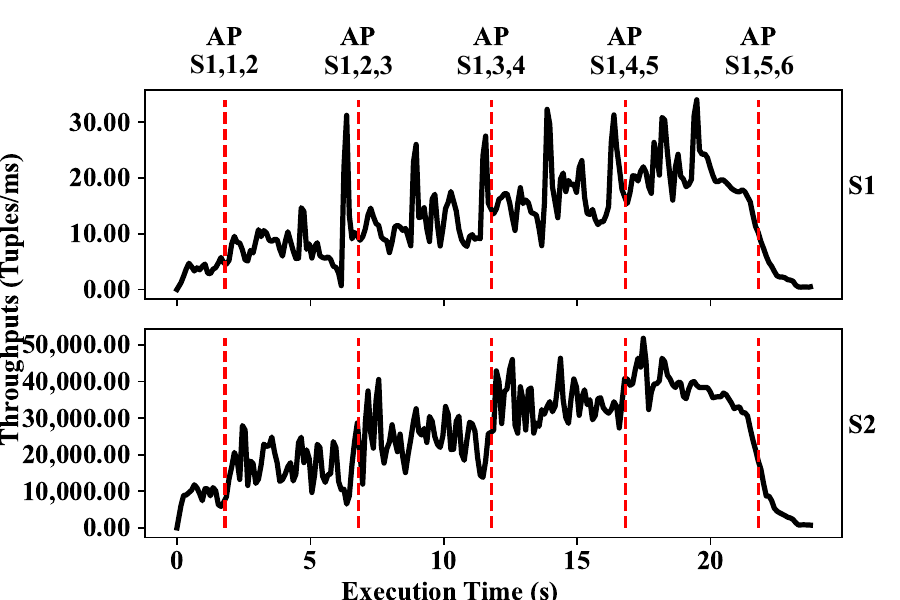}
  }
\\
\subfloat[Q5---Initial: 456ms; State transfer: \{S1: 3.56s; S2: 7.76s; S4: 2.11s\} \label{ISRS-Q5}]{
    \includegraphics[width=0.45\linewidth]{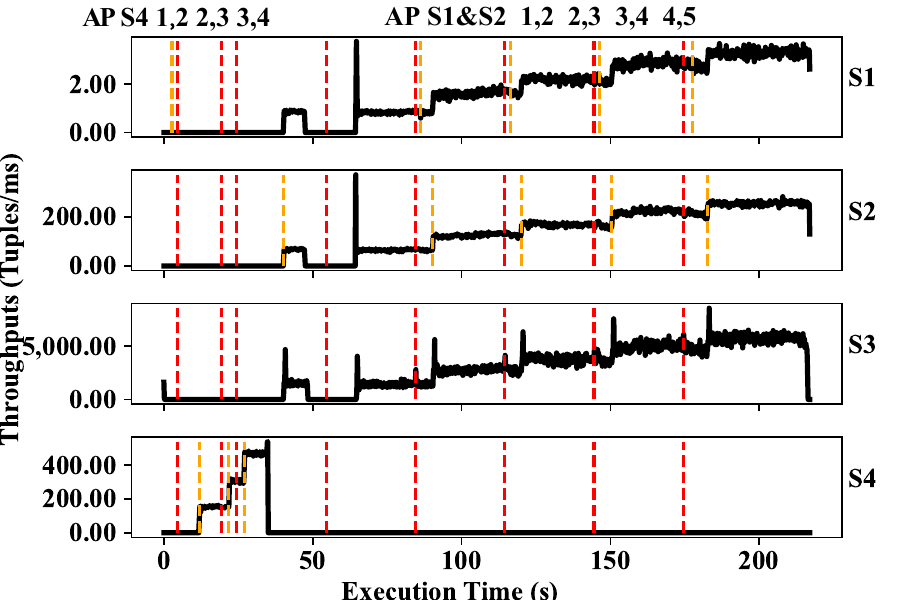}
  } 
 \subfloat[Q7---Initial: 468ms; State transfer: \{S1: 12.34s; S2: 14.76s; S7: 2.11s\} \label{ISRS-Q7}]{
    \includegraphics[width=0.45\linewidth]{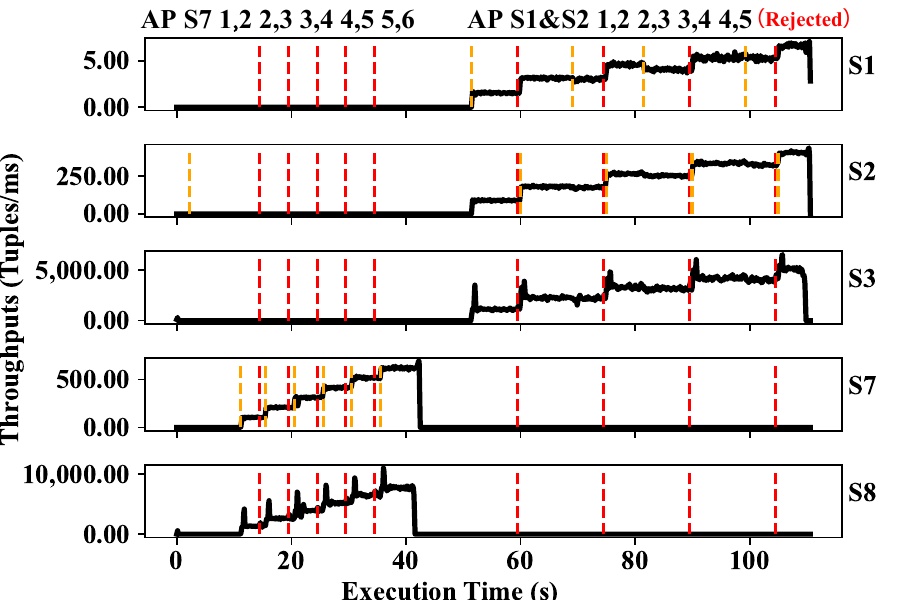}
  }
  \caption{Stage DOP tuning results \textnormal{-- Q1, Q3, Q5 and Q7.}}
  \label{multiDOPTuningResults}
\end{figure*}

\subsection{Stage DOP Runtime Tuning}
\label{ISRS_Exper}

In this section, we evaluate the intra-stage parallelism adjustment. We still use Q3 as an example. The initial intra-stage DOP of the Q3 is 1, the initial intra-task DOP is 1, and the intra-task DOP remains unchanged during execution. \cref{IntraStagetuning} illustrates the throughput variations for Q3 as stage parallelism is adjusted.

The notation ``AP S$n$, $a$,$b$'' denotes the adding parallelism for stage $n$ from $a$ to $b$ at the time marked by the red line. Initially, we adjusted the DOP for stage 3 three times, followed by five adjustments for stage 1. As both stages involve join operations, each parallelism adjustment necessitates hash table reconstruction, indicated by the yellow dashed lines appearing post-adjustment. The time interval $T_{build}$ between the red and yellow dashed lines reflects the duration for rebuilding the hash table, which depends on the data volume for the build side: 2.991s for stage 3 on average and 14.11 seconds for stage 1 on average. The last parallelism adjustment for stage 1 is rejected as the coordinator determines that the estimated remaining execution time is less than the stage's $T_{build}$. The overall execution time for the query is 194.76 seconds, achieving a 73.71\% reduction.

To evaluate the overhead of stage DOP tuning, we conducted a similar experiment to the one described in \cref{ITRS_Exper}, with results illustrated by the IntraStage-Inc curve in \cref{Q3ExeOrigin}. The overhead for stage DOP tuning includes hash table reconstruction in addition to task scheduling. Once tasks and drivers are created, the coordinator completes the scheduling process. Consequently, hash table reconstruction for multiple tasks occurs in parallel, enabling \sysname to efficiently add $n$ tasks simultaneously. The time required for hash table reconstruction is primarily divided into two components: data transfer (including shuffle and network transfer) and hash table construction. The larger the volume of data on the build side, the greater the interval between the IntraStage-Inc curve and the IntraStage curve becomes.

The query initialization time for Q3 and the state transfer time (i.e., the time from issuing a DOP adjustment request to completing the request) are provided in the caption of \cref{IntraStagetuning}. Additional experimental results for other queries is presented in \cref{multiDOPTuningResults}.

%\begin{figure}
%  \centering
 
%    \includegraphics[width=0.9\linewidth]{pics/Q3/Q3-IntraStage.pdf}
 
%  \caption{The stage throughput variation curves of intra-stage DOP tuning of Q3.}
%  \label{Q3IntraStagetuning}
%\end{figure}

\subsection{Partitioned Hash Join DOP Tuning}
\label{hash_Exper}

This section focuses on the evaluation of parallelism tuning for partitioned hash join. We use Q2J (\cref{2joinplan}) in \cref{ISRE} as an example for evaluation. The initial stage parallelism for query Q2J is set to 2, while the intra-task parallelism remains at 1 throughout execution. The execution time of the Q2J with the parallelism of 2 is 1331.991s.

\begin{figure}
  \centering
 
    \includegraphics[width=0.9\linewidth]{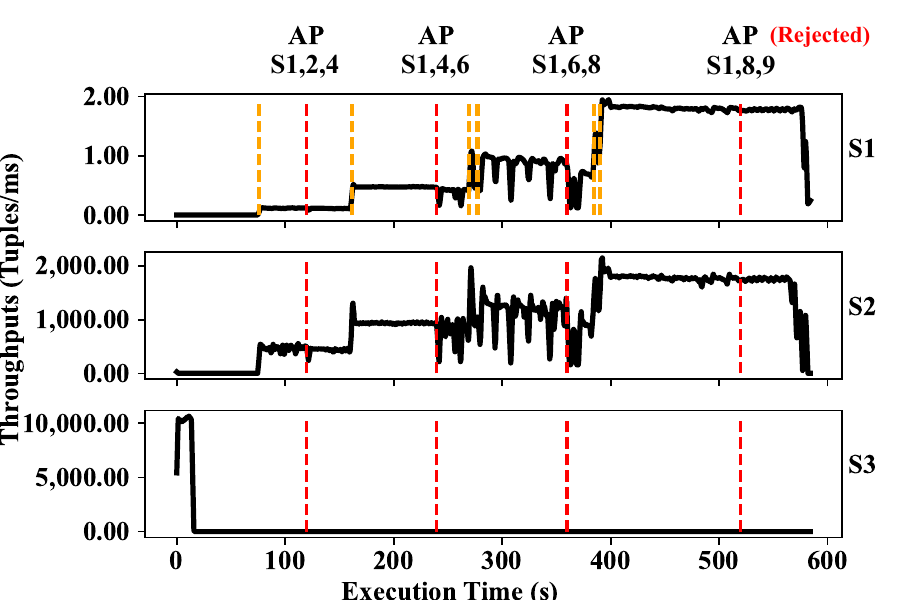}
 
  \caption{The stage throughput variation curves of the intra-stage parallelism tuning of Q2J.}
  \label{Q2JTuning}
\end{figure}

\subsubsection{DOP Switching Evaluation}

\cref{2joinplan} illustrates the distributed physical plan of Q2J, showing execution dependency between stage 1 and stage 3, and data dependency between stage 2 and stage 1. \cref{Q2JTuning} depicts the throughput variations during parallelism adjustments for Q2J. The query initialization time is 284ms. Stage DOP tuning takes an average of 23ms. The query's DOP is adjusted three times, with the last request rejected by the coordinator due to the remaining execution time being less than $T_{build}$. The notation ``AP S1,2,4'' indicates switching stage 1's parallelism from 2 to 4. The partitioned hash join requires the table reshuffling of the upstream stage and multiple hash table building of the current stage, resulting in multiple yellow dashed lines after each adjustment request.  We can see that the process of hash join is not interrupted during the process of hash table rebuilding. The total execution time for the query is 584.01 seconds, yielding a 56.16\% reduction in execution time. 

\begin{table}
  \caption{State transfer details of Q2J}
  \label{DOPswitchinglist}
  \begin{tabular}{cccc}
    \toprule
    DOP switching &Total time &Shuffle time & Build time\\
    \midrule
    \texttt{2 -> 4} & 42.67s & 12.55s & 30.12s\\
    \texttt{4 -> 6}& 29.03s & 8.80s & 21.03s\\
     \texttt{6 -> 8} & 21.61s & 5.12s & 16.49s\\
    \bottomrule
  \end{tabular}
\end{table}

In this query, the overhead of parallelism switching consists of shuffle time and hash table build time, the \cref{DOPswitchinglist} illustrates the details. For stages without partitioned hash joins, reducing parallelism requires only a few RESTful requests (tens to hundreds of milliseconds). In contrast, stages with partitioned hash joins always incur reshuffling when adjusting parallelism, but distributing data across more nodes can improve the DOP switching performance.

% DOP 2 to 4: Total time: 42.67s; Shuffle: 12.55s; Hash table build: 30.12s. DOP 4 to 6: Total time: 29.03s; Shuffle: 8.80s; Hash table build: 21.03s. DOP 6 to 8: Total time: 21.61s; Shuffle: 5.12s; Hash table build: 16.49s. 

%\begin{figure}
 %\centering
 
 %   \includegraphics[width=0.8\linewidth]{pics/Q2Join/Q2Join-Origin.pdf}
 
%  \caption{The stage throughput variation curves for Q2J with intra-stage parallelism of 2.}
%  \label{Q2JOrigin}
%\end{figure}

%\begin{figure}
%  \centering

% \includegraphics[width=0.65\linewidth]{pics/Q2Join/Q2Join-exe-origin.pdf}
  
%  \caption{The execution time of Q2J with different degrees of intra-stage parallelism.}
%  \label{Q2JExeOrigin}
%\end{figure}

\begin{figure}

\centering

\includegraphics[width=0.8\linewidth]{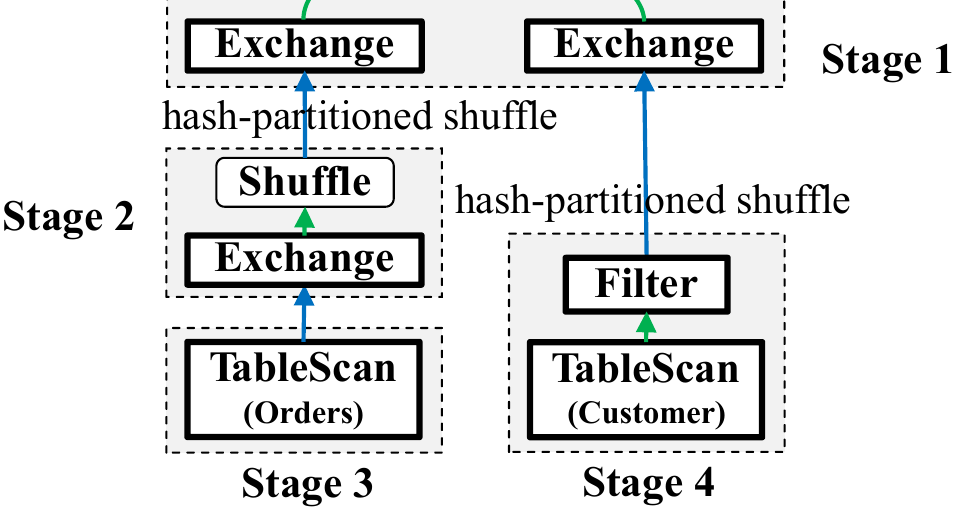}

\caption{The physical plan after adding shuffle stage.}
\label{shufflestageplan}
\end{figure}

%\begin{figure}
%\centering

%\includegraphics[width=0.9\linewidth]{pics/Q2Join/Q2JoinSO-4P-1coor-P6.pdf}

%\caption{The throughput curves of Q2J with no shuffle stage.}
%\label{shufflestageOrigin}
%\end{figure}

\begin{figure}
\centering

\includegraphics[width=0.9\linewidth]{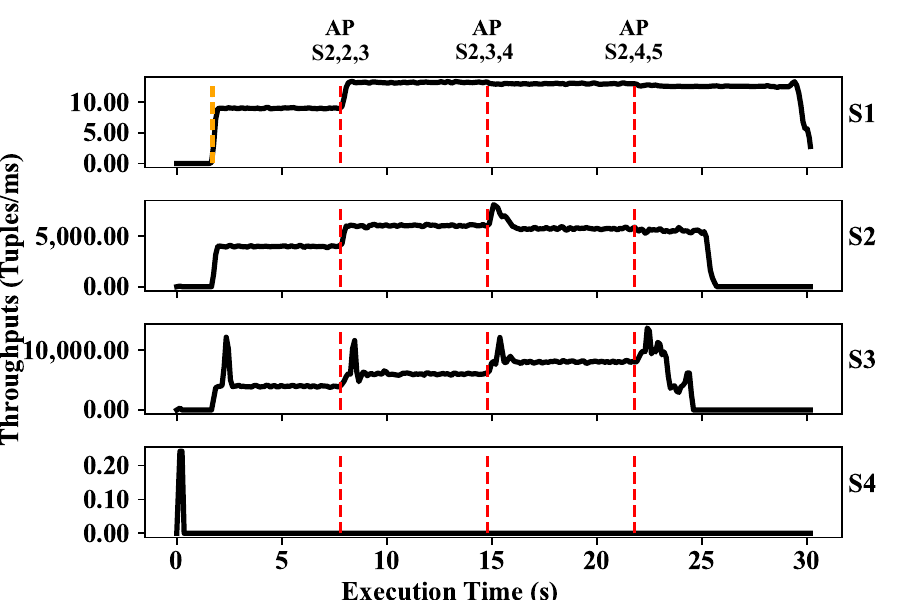}

\caption{The stage throughput variation curves of shuffle stage parallelism tuning.}
\label{shufflestageTuning}
\end{figure}

\subsubsection{Elastic Shuffle Stage Evaluation} Partitioned hash join presents two computational bottlenecks: shuffle bottlenecks and join bottlenecks. To evaluate the effectiveness of the shuffle stage, we used the query: "select count(o\_orderkey) from orders join customer on o\_custkey=c\_custkey where c\_ nationkey = 9" (the execution plan is similar to Q2J). Initially, the orders table was stored across two nodes to intentionally make the shuffle operation the query bottleneck. Executing the query under these conditions (S1 Stage DOP:10, Task DOP:1) resulted in a total execution time of 45.22 seconds. Next, as illustrated in \cref{shufflestageplan}, we added a shuffle stage downstream of the orders table and re-executed the query. The results in \cref{shufflestageTuning} show that the throughput of stages S1 and S3 gradually increased as the parallelism of stage S2 was increased. However, the effect of further parallelism increases became less significant because the query bottleneck shifted from the shuffle stage to the join stage. The query initialization time was 232 ms, and the parallelism switching overhead was 12 ms. The query's execution time was reduced to 30.21 seconds, representing a 33.19\% reduction in overall execution time.

\subsection{Automatic DOP Tuning}
\label{prediction}

\begin{figure}
  \centering
 
    \includegraphics[width=0.9\linewidth]{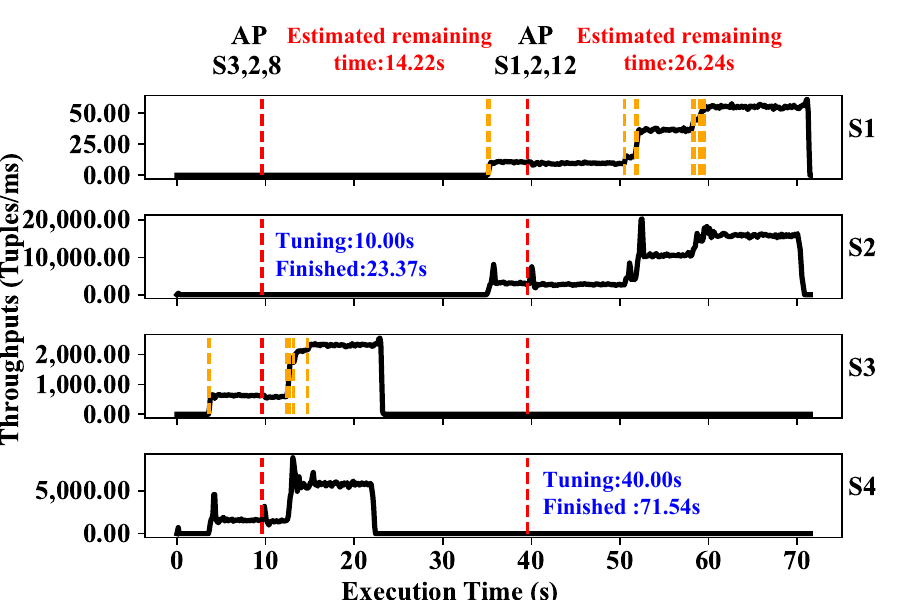}
 
  \caption{An Q3's stage DOP tuning throughput curves \textnormal{-- which marks the estimated time and the actual execution time.}}
  \label{Q3estimation}
\end{figure}

\begin{figure*}
  \centering
  \subfloat[Q2---Initial: 562ms; State transfer: \{S1: 14.34s; S10: 635ms\}. \label{autoTuneQ2}]{
    \includegraphics[width=0.45\linewidth]{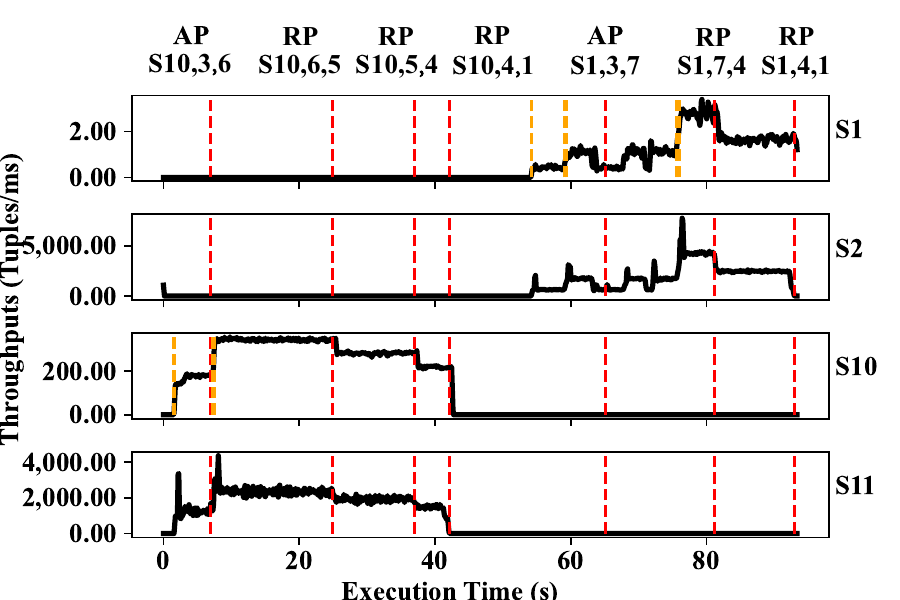}
  }
  \subfloat[Q3---Initial: 465ms; State transfer: \{S1: 13.65s; S3: 3.45s\}. \label{autoTuneQ3}]{
    \includegraphics[width=0.45\linewidth]{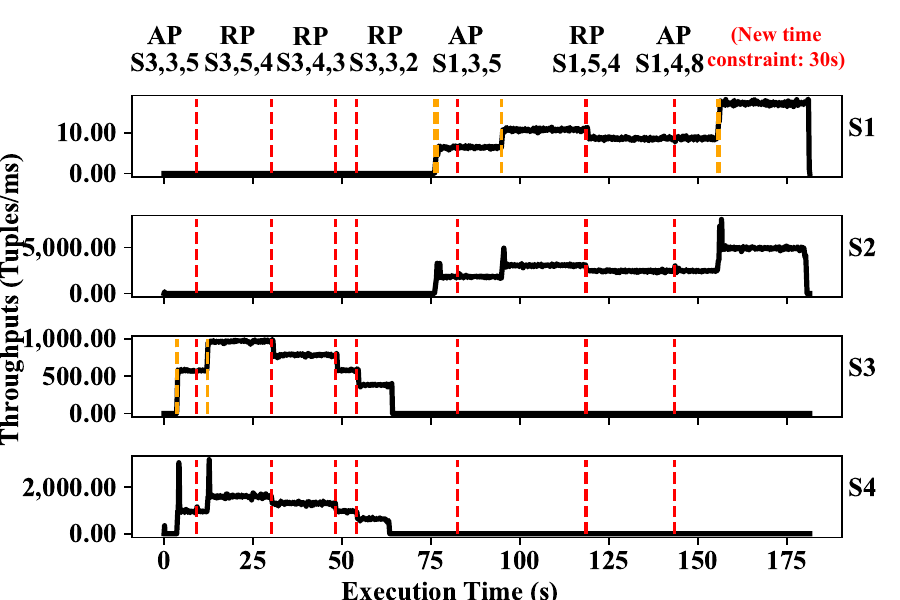}
  }

  \caption{Automatic DOP tuning throughput curves \textnormal{--  Q2 and Q3.}}
  \label{autoTuneResults}
\end{figure*}

In this section, we evaluate the effectiveness of the prediction of the stage remaining execution time and the effectiveness of the automatic DOP tuning.

\subsubsection{Stage Remaining Execution Time Prediction} \cref{Q3estimation} presents a throughput curve for stage DOP tuning in Q3. The query begins with a stage parallelism of 2 and a task parallelism of 3. Before each stage parallelism adjustment, the script executor estimates the remaining execution time and subsequently applies the DOP tuning request. For instance, before the first adjustment for stage 1, the prediction module calculates that changing the parallelism to 8 (2+6) results in a remaining execution time of 14.22 seconds. The estimation process is as follows: 1. The module first calculates the remaining execution time at the current parallelism as 59.28 seconds. 2. The hash table construction time is approximately 2.4s. 3. The estimated time is $(49.68-2.4)/4 + 2.4 = 14.22s$. The time point for the parallelism adjustment is at the 10th second, the time at the end of stage 3 is 23.37s, and the predicted time is 10+14.22=24.22s. In \cref{Q3estimation}, stage 1's parallelism adjustment occurs at the 40-second mark. The estimated completion time is 40+26.24s=66.24s. The actual finished time is at 71.55 seconds. The above data proves the accuracy of the time prediction of the predictor.

\subsubsection{DOP Auto-tuning} Below, we illustrate the impact of DOP auto-adjustment using queries Q2 and Q3 as examples.

The execution time of Q2 is primarily influenced by S1 (with upstream table scan stage S2) and S10 (with upstream table scan stage S11). For this auto-tuning task, the objective was to complete the query within 100 seconds. The DOP planning module initiated query with a stage DOP of 3 and a task DOP of 2, and it provided time constraints for each table scan stage, specifying that S11 should complete its table scan within 50 seconds and S2 within 50 seconds. The auto-tuning process is shown in \cref{autoTuneQ2}, where ``RP S$n$,$a$,$b$'' indicates that the auto-tuner reduced the parallelism of stage $n$ by from $a$ to $b$ at a specific time point. The only overhead incurred during parallelism reduction is the scheduling overhead, averaging 42 ms. As shown in \cref{autoTuneQ2}, the auto-tuner adjusts parallelism to meet time constraints while minimizing resource usage.

The execution time of Q3 is primarily determined by S1 (with upstream table scan stage S2) and S3 (with upstream table scan stage S4). In this task, the target was to complete the query within 200 seconds. The DOP planning module initiated query with a stage DOP of 3 and a task DOP of 2, and it set time constraints for S4 to complete its scan within 80 seconds and S2 within 120 seconds. The corresponding auto-tuning curves are provided in \cref{autoTuneQ3}. Unlike Q2, a new time constraint was introduced in real-time via the system UI at approximately the 150s, requiring S1 to finish execution within 30 seconds from that point. In response, the auto-tuner discarded the existing time-constrained plan and adjusted the DOP based on the updated constraint (AP S1,4,8). As shown in \cref{autoTuneQ3}, the auto-tuner successfully modified the DOP, enabling S1 to complete within the time constraints.

\section{Related Work}

\textbf{Intra-Query Elasticity}. Currently, the database and big data area mainly use ``dynamic query optimization'' to change resource usage during query execution. It can be categorized into three types: adaptive query processing, adaptive query execution, and query re-planning. Adaptive query processing \cite{judicious-reOp-adaptive-proecessing, re-opt2} is primarily applied in traditional standalone relational databases. These methods break down a query into multiple sub-queries, re-optimizing subsequent queries based on the results of earlier ones. Adaptive query execution \cite{runtime-dynamic-op, presto-cbo-hbo-adaptive, sparkAdaptiveExecutionWeb, redshift-scaling-adaptive} is more common in distributed environments, such as big data and cloud-native databases, and involves running queries in stages, using intermediate results to re-optimize the remaining query. Query re-planning focuses on adapting queries to new computing environments \cite{QoS-deadline-replan, QOOP} or execution configurations \cite{byteDance-stream-replan}, allowing re-planned queries to continue from a checkpoint. However, these methods typically require materializing intermediate results and halting data processing, making them unsuitable for frequent and efficient parallelism tuning.

\noindent\textbf{Inter-Query (Workload) Elasticity}. Current research in the field of cloud databases predominantly emphasizes the runtime elasticity of query workloads. These studies leverage the auto-scaling capabilities provided by cloud vendors to implement elastic computing. Prominent cloud databases, including Redshift \cite{redshift-scaling-adaptive,redshift2024}, Snowflake \cite{snowflakeWeb}, BigQuery \cite{bigQueryWeb}, and Azure SQL Database \cite{AzureWeb}, are well-equipped to efficiently support workload elasticity. In addition, serverless computing technologies \cite{cloudfunction1,cloudfunction2,cloudfunction3} enable users to execute computational tasks using cloud functions, offering a scalable and cost-effective alternative to traditional architectures. In this paper, we extend runtime elasticity research from inter-query to intra-query.

\noindent\textbf{Query optimization and scheduling of cloud databases}. Cloud databases primarily rely on rule-based and cost-based optimizers \cite{presto-cbo-hbo-adaptive, polardb-cbo, microsoft-rbo-workload, steering-ML-rbo-workload, basic-rbo, presto-rbo, scope-rbo, spark-rbo, bytegraph-cbo-rbo}. Various machine learning-based query optimization methods have been proposed \cite{ML-autowlm, ML-planOptimize-cuttlefish, ML-optimize, autosteer-bao-workload, bao2}. \cite{DOPML} uses machine learning to determine a near-optimal DOP for query execution. Most query schedulers \cite{ML-workload-umbra,schedule-quickstep} aim to optimize workloads. Additionally, numerous machine learning-based query schedulers have been developed \cite{workload-ML-bufferpool, ML-planOptimize-workload-skinnerdb, ML-workload-schedule-lsched, ML-workload-schedule-decima, ML-autowlm}.  However, these methods typically lack the capability for intra-query runtime optimization and scheduling.

\section{Conclusion and Future Work}

In this paper, we propose the concept of intra-query runtime elasticity, which enables a cloud-native OLAP engine to dynamically adjust the query degree of parallelism during execution. We introduce \sysname, the first IQRE query engine, capable of modifying parallelism at any point without pausing or interrupting the query execution. we experimentally demonstrate that Accordion is able to efficiently and automatically regulate the degree of parallelism to satisfy the user's query time constraints while minimizing computational resource usage. In the future, we will further enhance IQRE in three key directions: 1. Heterogeneous IQRE. Incorporating heterogeneous nodes, such as GPU nodes, to dynamically optimize query performance. 2. Dynamic execution plan. Modifying execution plans during query processing, such as inserting shuffle stage between stages. 3. Intelligent IQRE. Leveraging deep learning techniques to enable Accordion to better understand user preferences for query time and cost, allowing for more effective automatic selection and adjustment of DOP.

\balance
\bibliographystyle{ACM-Reference-Format}
\bibliography{mybib}

\end{document}